\documentclass[12pt,letterpaper]{article}
\usepackage{epsfig,rotating,setspace,latexsym,amsmath,epsf,amssymb,amsfonts,bm,theorem,cite,caption,subcaption,enumerate,longtable,accents}
\usepackage{algorithm,graphicx,epsf,authblk,epstopdf,url,color,multirow}
\usepackage{algpseudocode}
\newtheorem{theorem}{Theorem}

\newtheorem{remark}{Remark}
\newtheorem{lemma}{Lemma}
\newenvironment{Proof}[1]{\medskip\par\noindent{\bf Proof:\,}\,#1}{{\mbox{\,$\blacksquare$}\par}}

\newcommand{\bbE}{{\mathbb{E}}}

\allowdisplaybreaks

\begin{document}
	
\title{Timely Multi-Process Estimation Over Erasure Channels With and Without Feedback: Signal-Independent Policies
\thanks{This work was supported by the U.S. National Science Foundation under Grants CNS 21-14537 and ECCS 21-46099.}}

\author[1]{Karim Banawan}
\author[2]{Ahmed Arafa}
\author[3]{Karim G. Seddik} 
\affil[1]{\footnotesize Department of Electrical Engineering, Alexandria University, Egypt.}
\affil[2]{\footnotesize Department of Electrical and Computer Engineering, University of North Carolina at Charlotte, USA}
\affil[3]{\footnotesize Electronics and Communications Engineering Department, American University in Cairo, Egypt.}

\maketitle

\begin{abstract}
We consider a multi-process remote estimation system observing $K$ independent Ornstein-Uhlenbeck processes. In this system, a shared sensor samples the $K$ processes in such a way that the long-term average sum mean square error (MSE) is minimized using signal-independent sampling policies, in which sampling instances are chosen independently from the processes' values. The sensor operates under a total sampling frequency constraint $f_{\max}$. The samples from all processes consume random processing delays in a shared queue and then are transmitted over an erasure channel with probability $\epsilon$. We study two variants of the problem: first, when the samples are scheduled according to a Maximum-Age-First (MAF) policy, and the receiver provides an erasure status feedback; and second, when samples are scheduled according to a Round-Robin (RR) policy, when there is no erasure status feedback from the receiver. Aided by optimal structural results, we show that the optimal sampling policy for both settings, under some conditions, is a \emph{threshold policy}. We characterize the optimal threshold and the corresponding optimal long-term average sum MSE as a function of $K$, $f_{\max}$, $\epsilon$, and the statistical properties of the observed processes. Our results show that, with an exponentially distributed service rate, the optimal threshold $\tau^*$ increases as the number of processes $K$ increases, for both settings. Additionally, we show that the optimal threshold is an \emph{increasing} function of $\epsilon$ in the case of \emph{available} erasure status feedback, while it exhibits the \emph{opposite behavior}, i.e., $\tau^*$ is a \emph{decreasing} function of $\epsilon$, in the case of \emph{absent} erasure status feedback. 
\end{abstract}

\section{Introduction}

We study the problem of timely tracking of multiple random processes using shared resources. This setting arises in many practical situations of remote estimation and IoT applications. Recent works have drawn connections between the quality of the estimates at the destination, measured through mean square error (MSE), and the {\it age of information} (AoI) metric that assesses timeliness and freshness of the received data, see, e.g., the survey in \cite[Section~VI]{aoi-survey-jsac}. We extend these results to {\it multi-process} estimation settings in this work.

AoI is defined as the time elapsed since the latest received message has been generated at its source. It has been studied extensively in the literature in different contexts. This includes analyzing AoI in various queuing systems \cite{yates_age_1, ephremides_age_random, ephremides_age_management, ephremides_age_non_linear, yates-age-mltpl-src, talak-aoi-delay, inoue-aoi-general-formula-fcfs, soysal-aoi-gg11}, sampling and scheduling policy design for AoI optimization \cite{zou-waiting-aoi, modiano-age-bc, sun-age-mdp, zhou-age-iot, sun-cyr-aoi-non-linear}, minimizing AoI using energy harvesting and power constrained sensors \cite{tang-aoi-power-multi-state, yates_age_eh, jing-age-online, baknina-updt-info, arafa-age-online-finite, bacinoglu-aoi-eh-finite-gnrl-pnlty, leng-aoi-eh-cog-radio}, AoI analysis for multihop networks \cite{batu-aoi-multihop, bedewy-aoi-multihop}, AoI-optimal source coding design \cite{himanshu-age-source-coding}, fresh data market pricing using AoI \cite{zhang-arafa-aoi-pricing-wiopt}, cloud computing while maintaining data freshness \cite{arafa-aoi-compute}, and using AoI for scheduling in federated learning \cite{yang-arafa-aoi-fl}. 

Relevant to this work is the fact that AoI can be closely tied to MSE in random processes tracking applications. The works in \cite{klugel2019aoi-fr, mitra-estimation-graphs-aoi, chakravorty-estimation-pckt-drop-markov} characterize implicit and explicit relationships between MSE and AoI under different estimation contexts. References \cite{ayan-aoi-voi-cntrl, roth-mse-aoi-finite-blocklength}, however, consider the notion of the value of information (mainly through MSE) and show that optimizing it can be different from optimizing AoI. The quantification and optimality conditions of the value of information in feedback control systems of multi-dimensional Gaussian-Markov processes are further investigated in \cite{Soleymani-Baras-Hirche_VoI_quantification, Soleymani-Baras-Hirche-Johansson_VoI_optimality,soleymani_baras_johansson_2023}. Lossy source coding and distorted updates for AoI minimization is considered in \cite{ramirez-aoi-compression, bastopcu-aoi-distortion, bastopcu-partial-updates}. The notion of age of incorrect information (AoII) is introduced in in \cite{maatouk-aoii}, adding more context to AoI by capturing erroneous updates. The works in \cite{sun-wiener, OU_original} consider sampling of Wiener and Ornstein-Uhlenbeck (OU) processes for the purpose of remote estimation, and draw connections between MSE and AoI. Our recent work in \cite{sample_quantize} also focuses on characterizing the relationship of MSE and AoI, yet with the additional presence of coding and quantization. Reference \cite{guo2021optimal} shows the optimality of threshold policies for tracking OU processes under rate constraints.  The signal-dependent variant of the problem of remote estimation of OU processes is investigated in \cite{Rabi-Moustakides-Baras_adaptive_sampling} through casting the problem as choosing the optimal sequence of stopping times whenever the estimation error exceeds a designed envelope. The work \cite{Guo-Kostina_sampling_Markov} investigates optimal encoding and decoding policies to minimize MSE under rate constraint. The works \cite{Nourian-Leong-Dey_energy_Kalman,Nourian-Leong-Dey-Quevedo_optimal_transmission_Kalman} study the optimal transmission and energy allocation schemes for a sensor sending noisy measurements over a packet-dropping direct and feedback channels.

Reference \cite{OU_original} is closely related to our setting, in which optimal sampling methods to minimize the long-term average MSE for an OU process are derived. It is shown that if sampling times are independent of the instantaneous values of the process (signal-independent sampling) the minimum MSE (MMSE) reduces to an increasing function of AoI (age penalty). Then, threshold policies are shown optimal in this case, in which a new sample is acquired only if the expected age-penalty by the time it arrives surpasses a certain value. This paper extends \cite{OU_original} (and the related studies in \cite{sample_quantize,guo2021optimal}) to multiple OU processes.

\begin{figure}[t]
\center
\includegraphics[width=0.9\textwidth]{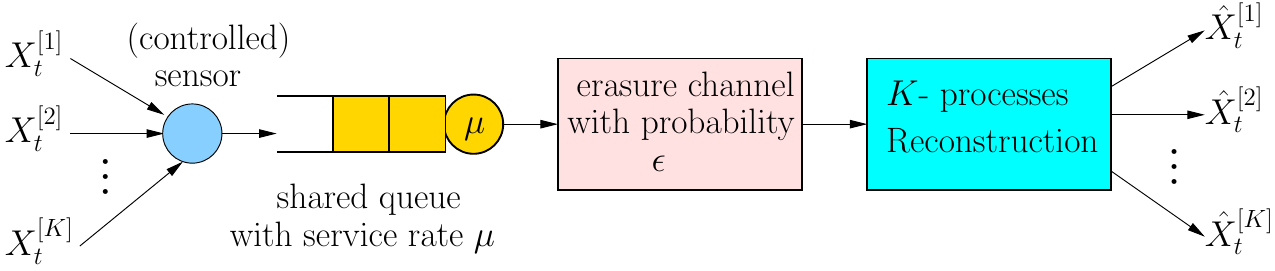}
\caption{System model.}
\label{fig:system model}
\end{figure}

In this paper, we study a remote sensing problem consisting of a shared controlled sensor, a shared queue, and a receiver (see Fig.~\ref{fig:system model}) to track $K$ independent, but not necessarily identical, OU processes.\footnote{The OU process is the continuous-time analog of the first-order autoregressive process \cite{ou-brownian-motion, doob-brownian-motion}, and is used to model various physical phenomena, and has relevant applications in control and finance.} The sensor transmits the collected samples over an erasure channel with probability $\epsilon$ after being processed for a random delay with service rate $\mu$. The sensor generates the samples \emph{at will}, subject to a total sampling frequency constraint $f_{\max}$. The goal is to minimize the long-term average sum MSE of the $K$ processes. We restrict our sampling policies to the family of \emph{signal-independent} policies, where the chosen sampling instants are identified based on the statistical measures of the processes and not the exact processes' values.\footnote{We note that there exists another family of sampling policies, namely, \emph{signal-dependent} policies. For this family of policies, the optimal sampling instants are defined as a function of the actual processes' values (or equivalently the instantaneous MSE of the reconstructed processes). Considering signal-dependent policies is an interesting future direction for this work, which is outside the scope of this paper.} 

In this work, we investigate two variants of the problem, which are different based on the \emph{availability of erasure status update} at the transmitter\footnote{We use the words sensor and transmitter interchangeably.}. In the first setting, the receiver sends an erasure status update to the transmitter. We focus on \emph{maximum-age-first} (MAF) scheduling, where the transmitter chooses the process with the largest AoI to be sampled. MAF scheduling results in obtaining a \emph{fresh} sample from the \emph{same} process until an unerased sample from that process is conveyed to the receiver. The erasure status update enables the use of MAF scheduling as the transmitter can accurately estimate the AoI of each process. In the second setting, the erasure status feedback is absent at the transmitter. In this setting, we focus on \emph{round-robin} (RR) scheduling, where the transmitter acquires samples from the processes in a \emph{fixed order} irrespective of the experienced erasure events. In both problem variants, we concentrate on \emph{stationary deterministic policies}, where the sampling policy needs to induce a stationary distribution across transmission epochs. Based on this, the optimal sampling policy reduces to optimizing a \emph{stationary waiting policy}. In this work, we do not aim to compare the optimal sum MSE with and without erasure feedback (naturally, a system with erasure feedback would prevail), rather we focus on the behavior of the optimal sampling policy.

We show that, for both problem settings, we can \emph{aggregate} the waiting times without affecting the sum MSE value, with a subtle difference between the two problem variants. Specifically, in the presence of erasure status feedback, the waiting times are aggregated at the beginning of the transmission \emph{epoch}, defined as the inter-delivery time between successful samples, which includes the \emph{successful} transmission of a sample from each process. In the absence of erasure status feedback, however, the transmitter aggregates the waiting times at the beginning of the transmission \emph{round}, which includes a transmission \emph{trial} of a sample from each process irrespective of the erasure outcome. We show that, for both problems, under MAF or RR scheduling, the optimal stationary deterministic policy is a \emph{threshold policy}. We characterize the optimal threshold $\tau^*(K,f_{\max},\epsilon,\boldsymbol{\theta},\boldsymbol{\sigma})$ and the corresponding long-term average sum MSE in terms of the processes' statistical properties ($\boldsymbol{\theta},\boldsymbol{\sigma}$), $\epsilon$, and $f_{\max}$. In both cases, the threshold is a maximum of two threshold values: one due to a nonbinding sampling frequency constraint scenario, and another due to a binding scenario. 

Surprisingly, our results show that \emph{optimal threshold $\tau^*$ behaves differently based on the availability erasure status feedback}. Specifically, our numerical results show that 1) the optimal threshold $\tau^*$ is an \emph{increasing} function in the erasure probability $\epsilon$ in the {presence} of erasure status feedback under MAF scheduling, 2) the optimal threshold $\tau^*$ is a \emph{decreasing} function in the erasure probability $\epsilon$ in the \emph{absence} of erasure status feedback under RR scheduling, and 3) the optimal threshold is an increasing function in the number of the observed processes $K$ for both problem variants under the respective scheduling policy.

\section{System Model}

We consider a sensing system in which $K$ independent, but not necessarily identical, OU processes are remotely monitored using a {\it shared} sensor that transmits samples from the processes over an erasure channel to a receiver. Denote the $k$th process value at time $s$ by $X_s^{[k]}$. Given $X_s^{[k]}$, the $k$th OU process evolves, for $t\geq s$, as \cite{ou-brownian-motion, doob-brownian-motion}
\begin{align} \label{eq_ou_evol}
X_t^{[k]}=X_s^{[k]}e^{-\theta_k(t-s)}+\frac{\sigma_k}{\sqrt{2\theta_k}}e^{-\theta_k(t-s)}W_{e^{2\theta_k(t-s)}-1},
\end{align} 
where $W_t$ denotes a Wiener process, while $\theta_k>0$ and $\sigma_k>0$ are fixed parameters that control how fast the process evolves. We study the system in steady-state, hence, we assume that the processes are initiated as $X_0^{[k]}\sim\mathcal{N}\left(0,\sigma_k^2/2\theta_k\right)$.\footnote{This way, the variance of $X_t^{[k]}$ is $\sigma_k^2/2\theta_k,~\forall t$, and the autocorrelation function is $\frac{\sigma_k^2}{2\theta_k}e^{-\theta_k|\tau|},~\tau\in\mathbb{R}$. Thus, large values of $\sigma_k$ or $\theta_k$ indicate a fast-varying process.}

To estimate $\left\{X_t^{[k]}\right\}$ at the receiver, the sensor acquires samples from the $k$th OU process at specific time instants $\left\{S_i^{[k]}\right\}$ and sends them to the receiver. Sampling instants are fully-controlled, i.e., samples are {\it generated-at-will}. We focus on \emph{signal-independent} sampling policies, in which the optimal sampling instants depend on the statistical measures of the processes and not on exact processes' values\footnote{We focus on the case in which the OU processes are non-observable prior to sampling. The case in which the processes are fully observable is to be studied in follow-up works.}.

The sensor must obey a {\it total sampling frequency constraint} $f_{\max}$. Let $\ell_i$ denote the $i$th sampling instant {\it regardless} of the identity of the process being sampled.  For example, if process $1$ is sampled two times consecutively and then process $2$ is sampled, then we have $\ell_1=S_1^{[1]}$, $\ell_2=S_2^{[1]}$, and $\ell_3=S_1^{[2]}$. Hence, it holds that $\left\{\ell_i\right\}_{i=1}^\infty\supseteq\cup_k\left\{S_i^{[k]}\right\}_{i=1}^{\infty}$. Therefore, the sampling constraint is expressed as follows:
\begin{align} \label{eq_smpl-const-genatwill}
    \liminf_{n\rightarrow\infty}\frac{1}{n}\mathbb{E}\left[\sum_{i=1}^n\ell_{i+1}-\ell_i\right]\geq\frac{1}{f_{\max}},
\end{align}
i.e., the long-term average inter-sampling time $\ell_{i+1}-\ell_i$ is constrained to be no smaller than $\frac{1}{f_{\max}}$, which indicates that the sensor shares the sampling budget $f_{\max}$ among the $K$ processes. Samples go through a shared processing queue, whose service model follows a Poisson process with service rate $\mu$, i.e., service times are independent and identically distributed (i.i.d.)~$\sim\exp(\mu)$ across samples. Served samples are prune to erasures with probability $\epsilon$, also occuring independently across samples. 

Samples are time-stamped prior to transmissions, and successfully-received samples from process $k$ determine the {\it age-of-information} (AoI) of that process at the receiver, denoted $\texttt{AoI}^{[k]}(t)$. AoI is defined as the time elapsed since the latest successfully received sample's time stamp. 

In this work, we investigate the effect of the presence/absence of the \emph{erasure status feedback} on specifying the sampling time instants $\left\{S_i^{[k]}\right\}$. Specifically, we study the following sampling scenarios:

\begin{enumerate}
    \item \emph{Erasure status feedback is available:} In this case, an immediate erasure status feedback is available at the sensor. We focus, in this case, on \emph{Maximum-Age-First (MAF) scheduling}, in which the processes are sampled according to their relative AoI's, with priority given to the process with highest AoI. Hence, at time $t$, process
	\begin{align}
	    \kappa(t)\triangleq\arg\max_k\texttt{AoI}^{[k]}(t)    
	\end{align}
is sampled. Observe that the value of $\kappa(t)$ will {\it not} change unless a successful transmission occurs. Therefore, in case of erasure events, a {\it fresh} sample is generated from the {\it same} process being served and transmission is re-attempted.

    \item \emph{Erasure status feedback is unavailable:} In this case, the receiver does not provide any feedback signaling to the sensor about the erasure status. In this case, MAF scheduling is not a viable scheduling policy as the sensor cannot identify the AoI of each process. Alternatively, we employ in this case the \emph{round-robin (RR) scheduling} with new samples, i.e., the sensor acquires a new sample from process~1, followed by a new sample from process~2, $\cdots$, etc., irrespective of the erasure events.  
      
\end{enumerate}

Although both MAF and RR policies schedule the processes' samples in the same order $1,2, \cdots, K$, the two scheduling policies differ in the erasure counter-measure (or lack thereof). Specifically, the MAF scheduling keeps re-attempting to sample the \emph{same} process until being successful. This is in contrast to RR scheduling, which \emph{keeps the order} of the processes regardless of the erasure events due to the absence of erasure status feedback.

Under both MAF and RR scheduling, and since the channel behaves similarly for all processes, each process will eventually be given an equal share of the allowed sampling budget, i.e., each process will be sampled at a rate of $f_{\max}/K$, and the sampling constraint in (\ref{eq_smpl-const-genatwill}) becomes
	\begin{align} \label{eq_smpl-const-maf}
        \liminf_{n\rightarrow\infty}\frac{1}{n}\mathbb{E}\left[\sum_{i=1}^nS_{i+1}^{[k]}-S_i^{[k]}\right]\geq\frac{K}{f_{\max}},\quad\forall k.
    \end{align}

Let $\tilde{S}_i^{[k]}$ denote the sampling instant of the $i$th {\it successfully}-received (unerased) sample from the $k$th process\footnote{In this work, we assume that a sample $S_i^{[k]}$ is either erased or exactly reconstructed without any distortion. This corresponds to the case when the samples are transmitted with infinite precision. Investigating the effects of quantization and coding is an interesting future direction that is outside the scope of this work. We refer the reader to \cite{sample_quantize}, where these effects are studied for single-process estimation problems.}, and (re-)define $S_i^{[k]}(m)$ as the sampling instant of the $m$th {\it attempt} to convey the $i$th sample of the $k$th process, $m=1,\dots,M_i^{[k]}$, with $M_i^{[k]}$ denoting the number of trials. Hence, we have $S_i^{[k]}(m)\leq\tilde{S}_i^{[k]},~\forall m$, with equality at $m=M_i^{[k]}$. Our channel model indicates that $M_i^{[k]}$'s are i.i.d.~$\sim \text{geometric}(1-\epsilon)$. Each sample $X_{S_i^{[k]}(m)}^{[k]}$ incurs a service time of $Y_i^{[k]}(m)$ time units with $Y_i^{[k]}(m)$'s being i.i.d.~$\sim\exp(\mu)$. The successfully-received sample, $X_{\tilde{S}_i^{[k]}}^{[k]}$, arrives at the receiver at time $D_i^{[k]}$, i.e.,
\begin{align}
D_i^{[k]}=\tilde{S}_i^{[k]}+Y_i^{[k]}\left(M_i^{[k]}\right).
\end{align}
Based on this notation, one can characterize the AoI of the $k$th OU process as follows:
\begin{align} \label{eq_aoi}
\texttt{AoI}^{[k]}(t)=t-\tilde{S}_i^{[k]}, \qquad D_{i}^{[k]} \leq t < D_{i+1}^{[k]}. 
\end{align}

The receiver collects the unerased samples from all processes and uses them to construct minimum mean square error (MMSE) estimates. Since the processes are independent, and by the strong Markov property of the OU process, the MMSE estimate for the $k$th process at time $t$, denoted $\hat{X}_t^{[k]}$, is based solely on the latest successfully-received sample from that process. Thus, for $D_{i}^{[k]} \leq t < D_{i+1}^{[k]}$, we have \cite{OU_original,sample_quantize}
\begin{align}
\hat{X}_t^{[k]}=&\mathbb{E}\left[X_t^{[k]}\Big|\tilde{S}_i^{[k]}, X_{\tilde{S}_i^{[k]}}\right] \nonumber \\
\stackrel{(\ref{eq_ou_evol})}{=}&X_{\tilde{S}_i^{[k]}}e^{-\theta_k(t-\tilde{S}_i^{[k]})}.
\end{align}
Let $\pi$ denote the scheduling policy, with $\pi\equiv\text{MAF}$ and $\pi\equiv\text{RR}$ in the presence and absence of erasure status feedback, respectively. Hence, the instantaneous mean square error (MSE) in estimating the $k$th process at time $t\in\left[D_i^{[k]},D_{i+1}^{[k]}\right)$ is \cite{OU_original,sample_quantize}
\begin{align}
\texttt{mse}_{\pi}^{[k]}\left(t,\tilde{S}_i^{[k]}\right)&\triangleq\bbE\left[\left(X_t^{[k]}-\hat{X}_t^{[k]}\right)^2\right] \\ 
&=\frac{\sigma_k^2}{2\theta_k}\left(1-e^{-2\theta_k\left(t-\tilde{S}_i^{[k]}\right)}\right), \label{MSE_value}
\end{align}
which is an increasing function of the AoI in (\ref{eq_aoi}) as we have $\texttt{mse}_{\pi}^{[k]}=\frac{\sigma_k^2}{2\theta_k}\left(1-e^{-2\theta_k\texttt{AoI}^{[k]}(t)}\right)$. Note that the MSE under MAF scheduling is different from that under RR scheduling, and hence the distinction using the subscript $\pi$. Next, we define the long-term time average MSE of the $k$th process as
\begin{align}\label{long_term_MSE}
\overline{\texttt{mse}_{\pi}^{[k]}}\!\triangleq\!\limsup_{T \rightarrow \infty}\frac{\sum_{i=1}^T \bbE\left[\int_{D_i^{[k]}}^{D_{i+1}^{[k]}} \texttt{mse}_{\pi}^{[k]}\left(t,\tilde{S}_i^{[k]}\right)dt\right]}{\sum_{i=1}^T\mathbb{E}\left[ D_{i+1}^{[k]}-D_i^{[k]}\right]}.
\end{align}  

Our goal is to choose the sampling instants to minimize a penalty function $g(\cdot)$ of $\left\{\overline{\texttt{mse}_\pi^{[k]}}\right\}$ under the sampling frequency constraint. More specifically, to solve the following problem in the presence/absence of erasure status feedback:
\begin{align} \label{opt_gen_maf}
\min_{\{S_i^{[k]}(m)\}} &\quad 	g\left(\overline{\texttt{mse}_{\pi}^{[1]}}, \cdots, \overline{\texttt{mse}_{\pi}^{[K]}}\right) \nonumber \\
\text{s.t.~~}  &~~ \liminf_{n\rightarrow\infty}\frac{1}{n}\mathbb{E}\left[\sum_{i=1}^nS_{i+1}^{[k]}-S_i^{[k]}\right]\geq\frac{K}{f_{\max}},~\forall k.
\end{align}

\section{Stationary Policies: Problem Re-Formulation}


In this section, we re-formulate problem \eqref{opt_gen_maf} in terms of a {\it stationary waiting policy} for each process. In the sequel, we provide the details of such re-formulation for both cases of erasure feedback availability.

\subsection{Problem Re-Formulation in the Presence of Erasure Feedback}\label{sec: reformulation with feedback}

\begin{figure}[t]
\center
\includegraphics[width=1\textwidth]{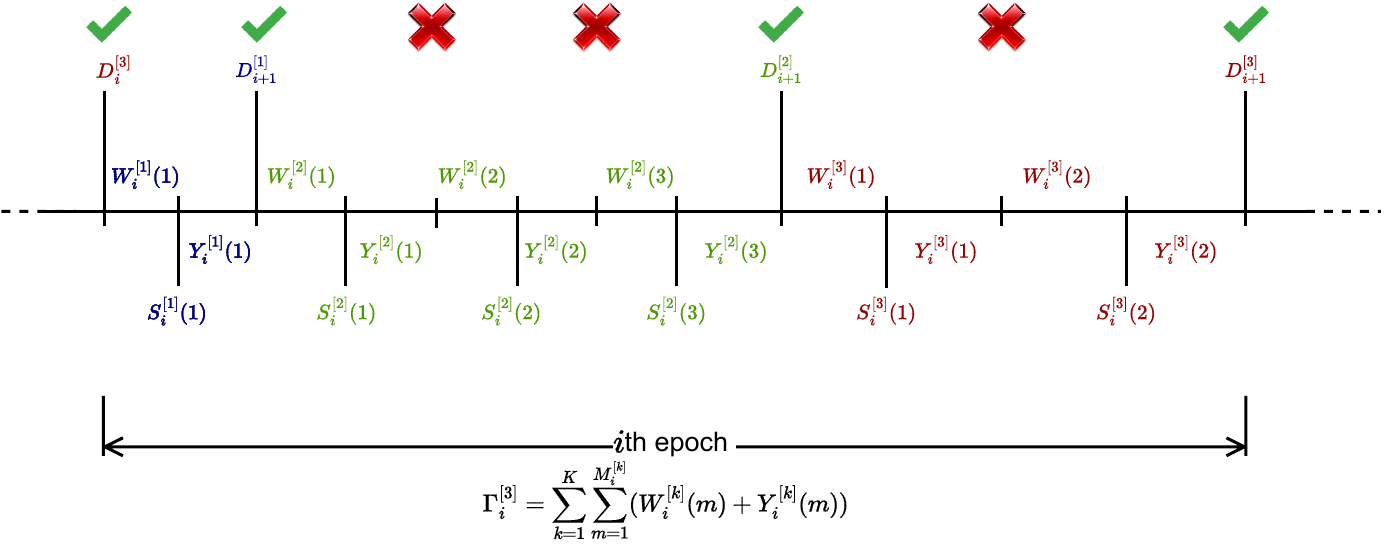}
\caption{Timing diagram of a typical epoch for $K=3$ processes in the presence of erasure feedback and MAF scheduling. In this example, process~1 (in blue) is successful after one trial, process~2 (in green) is successful after 3 trials, and process~3 (in red) is successful after 2 trials.}
\label{fig: FB_epoch}
\end{figure}

In this subsection, we focus on the case in which the erasure status feedback is available. We define $W_i^{[k]}(m)$ as the $m$th waiting time before the $m$th transmission attempt towards conveying the $i$th sample from the $k$th process, $1\leq m\leq M_i^{[k]}$. Without loss of generality, let the MAF schedule be in the order $1,2,\dots,K$. Thus, we have (see Fig.~\ref{fig: FB_epoch})
\begin{align}
    S_i^{[k]}(m)=D_{i}^{[k-1]}&+\sum_{j=1}^{m-1} Y_i^{[k]}(j)+\sum_{j=1}^{m}W_i^{[k]}(j),
\end{align}
with $D_i^{[0]}\triangleq D_{i-1}^{[K]}$. Problem \eqref{opt_gen_maf} now reduces to optimizing the waiting times $\left\{W_i^{[k]}(m)\right\}$. We now define the $i$th \emph{epoch} of the $k$th process, denoted $\Gamma_i^{[k]}$, as the inter-reception time in between its $i$th and $(i+1)$th {\it unerased} samples, i.e.,
\begin{align}
    \Gamma_i^{[k]}=D_{i+1}^{[k]}-D_i^{[k]}.
\end{align}

In this work, we focus on {\it stationary waiting policies} in which the waiting policy $\left\{W_i^{[k]}(m)\right\}$ has the same distribution across all processes' epochs. Note that under MAF scheduling, each process epoch entails a successful transmission of every other process. This, together with the fact that service times and erasure events are i.i.d., induces a stationary distribution across all processes' epochs. Therefore, dropping the indices $i$ and $k$, we have $\Gamma_i^{[k]}\sim\Gamma,~\forall i,k$, where
\begin{align}\label{epoch_presence_feedback}
\Gamma=\sum_{k=1}^K \sum_{m=1}^{M^{[k]}} W^{[k]}(m) + Y^{[k]}(m).
\end{align}
Now consider a typical epoch for the $k$th process. By stationarity, one can write \eqref{long_term_MSE} as
\begin{align}\label{mse_Gamma}
    \overline{\texttt{mse}_{\text{MAF}}^{[k]}}=\frac{ \bbE\left[\int_{D^{[k]}}^{D^{[k]}+\Gamma} \texttt{mse}_{\text{MAF}}^{[k]}\left(t,\tilde{S}^{[k]}\right)dt\right]}{\mathbb{E}\left[\Gamma \right]}.
\end{align}
where $D_i^{[k]}\sim D^{[k]}$ and $\tilde{S}_i^{[k]}\sim \tilde{S}^{[k]}$, $\forall i$. In the sequel, {\it we treat the $K$th (last) process's epoch as the typical epoch.} 

\begin{figure}[t]
\center
\includegraphics[width=1\textwidth]{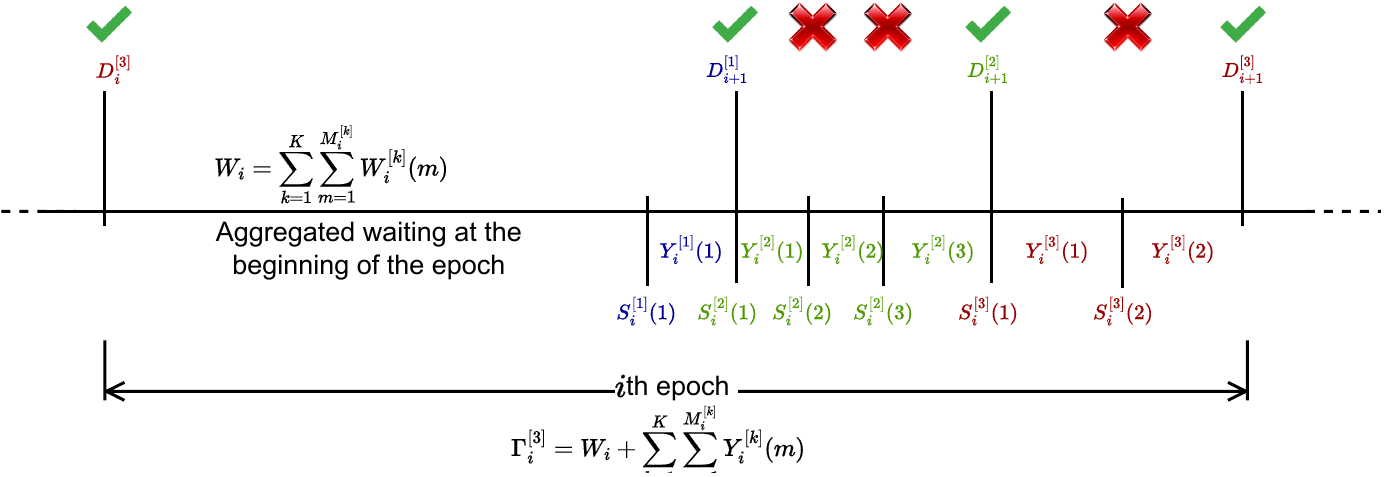}
\caption{Illustration of Lemma~\ref{lemma: reformulation presence of feedback}. The waiting times are grouped to the beginning of the epoch without affecting the long-term average MSE penalty.}
\label{fig: FB_epoch_aggregation}
\end{figure}

In the next lemma, we prove an important structural result, which asserts that the positions of the waiting times do not matter. Specifically, we show that one can achieve the same long-term average MSE penalty by {\it grouping all the waiting times at the beginning of the (typical) epoch} (see Fig.~\ref{fig: FB_epoch_aggregation}).

\begin{lemma}\label{lemma: reformulation presence of feedback}
Under signal-independent sampling with MAF scheduling, problem \eqref{opt_gen_maf} is equivalent to the following optimization problem for stationary waiting policies:
\begin{align} \label{opt_gen_maf_onewait}
\min_{W \geq 0} &\qquad 	g\left(\overline{\texttt{mse}_{\text{MAF}}^{[1]}}, \cdots, \overline{\texttt{mse}_{\text{MAF}}^{[K]}}\right) \nonumber \\
\mbox{s.t.} &\qquad \bbE\left[(1-\epsilon)W+\sum_{k=1}^{K}Y^{[k]}\right] \geq \frac{K}{f_{\max}},
\end{align}
where $W\triangleq\sum_{k=1}^K \sum_{m=1}^{M^{[k]}} W^{[k]}(m)$ and the waiting is only performed at the \emph{beginning} of the epoch.
\end{lemma}

\begin{Proof}
By inspection of the average MSE function in \eqref{mse_Gamma}, since $\Gamma=\sum_{k=1}^K \sum_{m=1}^{M^{[k]}} W^{[k]}(m)+\sum_{k=1}^K \sum_{m=1}^{M^{[k]}} Y^{[k]}(m)$, the waiting times appear in the numerator and denominator as the sum
\begin{align}
\sum_{k=1}^K \sum_{m=1}^{M^{[k]}} W^{[k]}(m).
\end{align}
Thus, for the optimal waiting times $\left\{W^{[k]^*}(m)\right\}$ that solve the optimization problem in \eqref{opt_gen_maf}, the waiting time $W^*=\sum_{k=1}^K\sum_{m=1}^{M^{[k]}} W^{[k]^*}(m)$ achieves the same $\overline{\texttt{mse}_{\text{MAF}}^{[k]}}$. Conversely, starting with $W^*$ in the objective function of (\ref{opt_gen_maf_onewait}) and breaking it arbitrarily to any waiting times such that $W^*=\sum_{k=1}^K \sum_{m=1}^{M^{[k]}} W^{[k]^*}(m)$ gives the same objective function in \eqref{opt_gen_maf}.

As for the sampling constraint, by observing the telescoping sum in \eqref{eq_smpl-const-maf}, we have that for process $k$,
\begin{align}\label{sampling_constraint}
    \liminf_{n\rightarrow \infty} \frac{1}{n} \bbE\left[\sum_{i=1}^n S_{i+1}^{[k]}-S_i^{[k]}\right]
    &=\liminf_{n\rightarrow \infty} \frac{1}{n} \bbE\left[S_{n+1}^{[k]}\right].
\end{align}
Now define $e(n)$ to be the index of the epoch corresponding to the $n$th successfully-received sample. With $W=\sum_{k=1}^K\sum_{m=1}^{M^{[k]}} W^{[k]^*}(m)$, we can write the sampling constraint as
\begin{align}
    \liminf_{n \rightarrow \infty} \frac{e(n)}{n}&\cdot \frac{\bbE[S_{n+1}^{[k]}]}{e(n)} \nonumber \\
    =&\frac{1}{\bbE[M^{[k]}]} \cdot \liminf_{n \rightarrow \infty} \frac{1}{e(n)} \Bigg(\sum_{i=1}^{e(n)-1}\bbE\Bigg[\sum_{k=1}^K  \sum_{m=1}^{M_i^{[k]}}W_i^{[k]}(m)+Y_i^{[k]}(m)\Bigg]+o\left(e(n)\right)\Bigg)\label{SLLN_M}\\
    =&\frac{1}{\bbE[M^{[k]}]} \cdot\liminf_{n \rightarrow \infty} \frac{1}{e(n)}\!\!  \sum_{i=1}^{e(n)-1}\left( \bbE[W]+ \bbE\left[M_i^{[k]}\right]\cdot \bbE\left[\sum_{k=1}^KY_i^{[k]}\right]\right) \label{wald_sampling}\\
    =&\bbE\left[(1-\epsilon)W+\sum_{k=1}^K Y^{[k]}\right],
\end{align}
where \eqref{SLLN_M} follows from the strong law of large numbers and the fact that the time spent in the $e(n)$th epoch, $\Delta=\sum_{\tilde{k}=1}^{k-1}\sum_{m=1}^{M^{[\tilde{k}]}} W_i^{[\tilde{k}]}(m)+Y_i^{[\tilde{k}]}(m)+\sum_{m=1}^{\tilde{m}} W_i^{[k]}(m)+Y_i^{[k]}(m)$, is $o(e(n))$ and hence $\liminf_{n \rightarrow \infty} \frac{\Delta}{e(n)} =0$; equation \eqref{wald_sampling} follows from Wald's identity. 
\end{Proof}

We now have the following remark:

\begin{remark} \label{rmrk_smpl}
Observe that the sampling constraint in problem (\ref{opt_gen_maf_onewait}) will not be active if $f_{\max}>\mu$. This is intuitive since the inter-sampling time, on average, would be larger than the minimum allowable sampling time, controlled by the maximum allowable sampling frequency, in this case. 

If the sampling constraint is binding, which occurs only if $f_{\max}<\mu$, the average waiting time would monotonically increase with the erasure probability $\epsilon$. This is true because no waiting is allowed in between unsuccessful transmissions, whose rate increases with $\epsilon$. Hence, to account for the expected large number of back-to-back sample transmissions in the epoch, one has to wait for a relatively larger amount of time at its beginning so that the sampling constraint is satisfied.
\end{remark}

\subsection{Problem Re-Formulation in the Absence of Erasure Feedback}

In this subsection, we turn our attention to the other case, in which the erasure status feedback is not available. Throughout our treatment, we highlight the main differences compared to the re-formulation in Section~\ref{sec: reformulation with feedback}.

Similar to Section~\ref{sec: reformulation with feedback}, we define the $i$th epoch of the $k$th process $\Gamma_i^{[k]}$ as the inter-reception time in between its $i$th and $(i+1)$th successfully received samples. Despite the identical definition, we note that the epoch $\Gamma_i^{[k]}$, in this case, consists of multiple \emph{transmission rounds}. Each round implies transmitting a \emph{new sample} from each process in a round-robin fashion, i.e., in the order of $1, 2, 3, \cdots, K$ irrespective of the erasure outcome of transmission at the receiver. 

\begin{figure}[t]
\center
\includegraphics[width=1\textwidth]{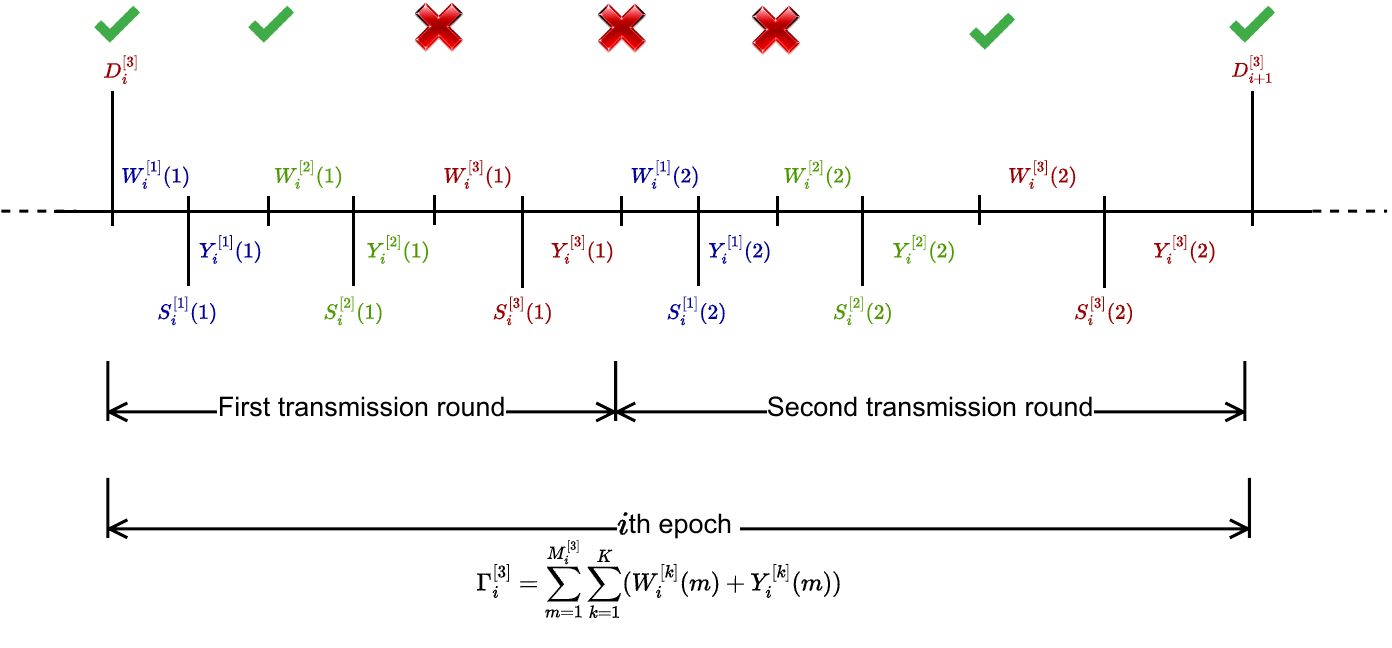}
\caption{Timing diagram of a typical epoch for $K=3$ processes in the absence of erasure feedback and RR scheduling. In this example, we focus on characterizing the epoch of process~3, which is successful after 2 transmission rounds. The erasures experienced by the remaining of the processes is irrelevant to the epoch of process~3.}
\label{fig: no_FB_epoch}
\end{figure}

The transmitter introduces stationary waiting times $\{W_i^{[k]}(m)\}$ between successive samples in each transmission round, i.e., the transmitter waits for $W_i^{[k]}(m)$ time units after transmitting a sample from the $(k-1)$th process before taking a new sample from the $k$th process at the $m$th transmission round (see Fig.~\ref{fig: no_FB_epoch}). With a slight abuse of notation, let $Y_i^{[k]}(m)$ be the service time of the sample from the $k$th process in the $i$th epoch at the $m$th transmission round. We have that $Y_i^{[k]}(m)$'s are i.i.d. such that $Y_i^{[k]}(m) \sim \text{exp}(\mu)$. Let $M_i^{[k]}$ denote the total number of transmission rounds needed to convey the $i$th sample of the $k$th process. Consequently, $M_i^{[k]} \sim \text{geometric}(1-\epsilon)$. We note that $M_i^{[k]}$ controls the epoch length $\Gamma_i^{[k]}$ irrespective of the other processes' erasure statuses. Without loss of generality, we focus on the $K$th process. Under RR scheduling, we can express the sampling instant $S_i^{[K]}(m)$, for $m=1,\cdots, M_i^{[K]}$ as
\begin{align}
    S_i^{[K]}(m)=S_i^{[K]}(m-1)+Y_i^{[K]}(m-1)+\sum_{k=1}^{K-1} Y_i^{[k]}+\sum_{k=1}^{[K]} W_i^{[k]}(m).
\end{align}

Based on the aforementioned description, we can write the $i$th epoch length corresponding to the $k$th process as
\begin{align}
    \Gamma_i^{[k]}=\sum_{m=1}^{M_i^{[k]}}\sum_{k=1}^{K} W_i^{[k]}(m)+Y_i^{[k]}(m).
\end{align}

By imposing the stationarity restriction of the waiting policy, the i.i.d. statistics of the service times and the erasure events, and the fact that epoch corresponding to the $k$th process entails repeating the same cycle of transmissions $M_i^{[k]} \sim \text{geometric}(1-\epsilon)$ times, we can drop the indices $i$ and $k$ as we have done in Section~\ref{sec: reformulation with feedback} to have\footnote{It is worth noting that the epoch definitions in \eqref{epoch_presence_feedback} and \eqref{epoch_absence_feedback} differ only in the order of summation. This signifies the fact that in the presence of erasure status feedback, the transmitter takes $M_i^{[k]}$ samples from the $k$th process until being successful before sampling the $(k+1)$th process. This is in contrast to passing by all processes \emph{once} and repeating this cycle $M$ times to complete the epoch in the absence of erasure status feedback.}
\begin{align}\label{epoch_absence_feedback}
    \Gamma=\sum_{m=1}^{M}\sum_{k=1}^{K} W^{[k]}(m)+Y^{[k]}(m)
\end{align}

Consequently, despite the subtle differences and the slight abuse of notation, we can still write the long-term average MSE of the $k$th process as that in \eqref{mse_Gamma} (after replacing $\text{MAF}$ with $\text{RR}$).  

Now, similar to Lemma~\ref{lemma: reformulation presence of feedback}, we can aggregate all waiting times in the $m$th round at the \emph{beginning of each transmission round}, i.e., the transmitter waits for
\begin{align}
W_i(m)=\sum_{k=1}^{K} W_i^{[k]}(m)
\end{align}
time units in the $m$th round. Specifically, the transmitter waits for $W_i(1)=\sum_{k=1}^{K} W_i^{[k]}(1)$  at the first round, after which the transmitter acquires and transmits a sample from process~1, followed by a sample from process~2, $\cdots$, followed by a sample from process~$K$, without any waiting times in between. Then the transmitter waits for $W_i(2)=\sum_{k=1}^{K} W_i^{[k]}(2)$ before completing the second round-robin cycle (second round) of sampling across all processes, $\cdots$, etc. Similar to Lemma~\ref{lemma: reformulation presence of feedback}, this aggregation does not affect the long-term average MSE and results in a simpler equivalent optimization problem (see Fig.~\ref{fig: no_FB_epoch_aggregation}). This is summarized in the following lemma:

\begin{figure}[t]
\center
\includegraphics[width=1\textwidth]{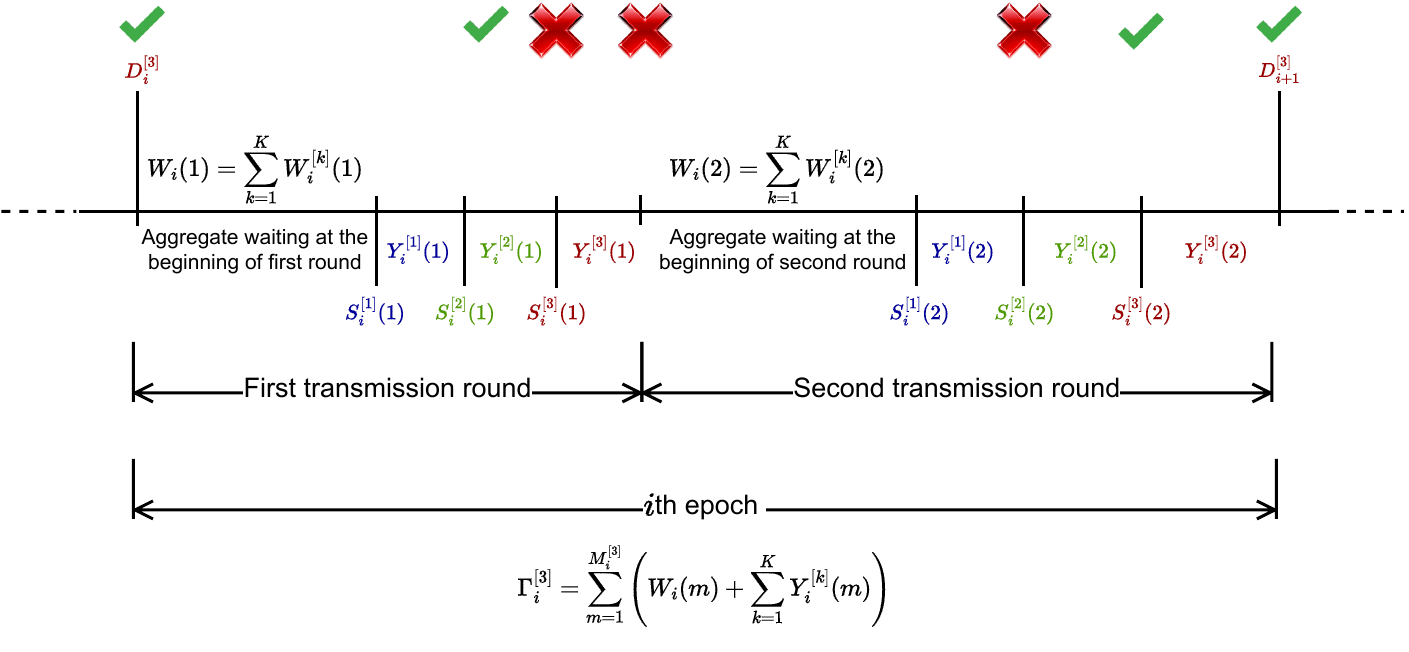}
\caption{Illustration of Lemma~\ref{lemma: reformulation absence of feedback}. The waiting times are grouped to the beginning of each transmission round without affecting the long-term average sum MSE.}
\label{fig: no_FB_epoch_aggregation}
\end{figure}

\begin{lemma}\label{lemma: reformulation absence of feedback}
    Under signal-independent sampling with RR scheduling, problem \eqref{opt_gen_maf} is equivalent to the following optimization problem for stationary waiting policies:
\begin{align} \label{opt_gen_RR_onewait}
\min_{W \geq 0} &\qquad 	g\left(\overline{\texttt{mse}_{\text{RR}}^{[1]}}, \cdots, \overline{\texttt{mse}_{\text{RR}}^{[K]}}\right) \nonumber \\
\mbox{s.t.} &\qquad \bbE\left[W(m)+\sum_{k=1}^{K}Y^{[k]}\right] \geq \frac{K}{f_{\max}},
\end{align}
where $W(m)\triangleq\sum_{k=1}^K W^{[k]}(m)$ and the waiting is only performed at the beginning of the $m$th transmission round.
\end{lemma}

The proof of Lemma~\ref{lemma: reformulation absence of feedback} follows the exact steps of the proof of Lemma~\ref{lemma: reformulation presence of feedback} while noting that the waiting time here is at the beginning of  \emph{every transmission round} and not at the beginning of the epoch as in Lemma.~\ref{lemma: reformulation presence of feedback}. Hence, the proof is removed to avoid unnecessary repetitions.

\begin{remark}
    We note that, in the absence of erasure status feedback, the transmitter cannot aggregate all waiting times at the beginning of the epoch. This is due to the fact that the epoch beginning time $D_{i}^{[k]}$ and end time $D_{i+1}^{[k]}$ are only visible at the receiver side due to the absence of the erasure status feedback. 
\end{remark}

\begin{remark}
    The sampling constraint in \eqref{opt_gen_RR_onewait} lacks the erasure effect in \eqref{opt_gen_maf_onewait}. This is due to the fact that the number of samples at each transmission round is $K$ samples, which is independent of the erasure events. This is in contrast to the number of samples in an epoch in \eqref{opt_gen_maf_onewait}, which is significantly dependent on the erasure probability. 
\end{remark}

\begin{remark}
    We note that when $\epsilon=0$, $\text{MAF}\equiv\text{RR}$ scheduling, and the optimization problems \eqref{opt_gen_maf} with and without erasure status feedback are indeed the same (i.e., problems \eqref{opt_gen_maf_onewait} and \eqref{opt_gen_RR_onewait} are the same) as the transmitter acquires one sample only from each process during any epoch.
\end{remark}


We discuss optimal solutions of problems \eqref{opt_gen_maf_onewait} and \eqref{opt_gen_RR_onewait} over the next two sections, respectively, for specific MSE penalty functions.

\section{Optimal Waiting Threshold and Minimum Sum MSE Characterization with Erasure Feedback}\label{sec: solution presence feedback}

In this section, we provide the optimal solution of problem (\ref{opt_gen_maf_onewait}) for a {\it sum MSE} penalty
\begin{align}\label{sum_mse}
	g\left(\overline{\texttt{mse}_{\text{MAF}}^{[1]}}, \cdots, \overline{\texttt{mse}_{\text{MAF}}^{[K]}}\right)=\sum_{k=1}^K \overline{\texttt{mse}_{\text{MAF}}^{[k]}},
\end{align}
together with a {\it stationary deterministic} waiting policy, in which the waiting value at the beginning of an epoch is given by a deterministic function $w(\cdot)$ of the previous epoch's total service time, denoted $\tilde{Y}$, i.e., 
\begin{align}
\tilde{Y}\sim\sum_{k=1}^{K}\sum_{m=1}^{M^{[k]}} Y^{[k]}(m).
\end{align}
Note that such choice of waiting policies emerges naturally since the MSE is an increasing function of the AoI, whose value at the start of the epoch is, in turn, an increasing function of $\tilde{Y}$. Stationary deterministic policies have been used extensively in similar contexts in the literature, see, e.g., \cite{sun-wiener, OU_original, sample_quantize}, and have been shown to perform optimally. 

Formally, substituting the above into problem (\ref{opt_gen_maf_onewait}), we now aim at solving the following functional optimization problem:
\begin{align}\label{sumMSE}
    \min_{w(\cdot)\geq 0} &\quad \frac{ \sum_{k=1}^K \bbE\left[\int_{D^{[k]}}^{D^{[k]}+\Gamma} \texttt{mse}_{\text{MAF}}^{[k]}\left(t,\tilde{S}^{[k]}\right)dt\right]}{\mathbb{E}\left[\Gamma \right]} \notag\\
    \text{s.t.~} &\quad \bbE\left[w\left(\tilde{Y}\right)\right] \geq \frac{1}{1-\epsilon}\left(\frac{K}{f_{\max}}-\frac{K}{\mu}\right).
\end{align}

Theorem~\ref{thm1} below provides the optimal solution of problem \eqref{sumMSE}. We use the compact vector notation $\boldsymbol{\theta}\triangleq[\theta_1 \:\: \theta_2 \: \cdots \: \theta_K]$ and $\boldsymbol{\sigma}\triangleq[\sigma_1^2 \:\: \sigma_2^2 \: \cdots \: \sigma_K^2]$.

\begin{theorem}\label{thm1}
The optimal waiting policy $w^*(\cdot)$ that solves problem (\ref{sumMSE}) is given by the \emph{threshold} policy
\begin{align}\label{waiting-policy-opt}
    w^*(z)=\left[\tau^*_{\text{MAF}}(K,f_{\max},\epsilon,\boldsymbol{\theta},\boldsymbol{\sigma})-z\right]^+,
\end{align}
where the optimal threshold $\tau^*_{\text{MAF}}(K,f_{\max},\epsilon,\boldsymbol{\theta},\boldsymbol{\sigma})$ is given by
\begin{align}\label{threshold}
  \tau^*_{\text{MAF}}=\max\left\{G_{\boldsymbol{\theta},\boldsymbol{\sigma}}^{-1}\left(\beta^*\right), H^{-1}\left(\frac{1}{(1-\epsilon)}\left[ \frac{K}{f_{\max}}-\frac{K}{\mu}\right]^+\right)\right\},
\end{align}
in which
\begin{align}
G_{\boldsymbol{\theta},\boldsymbol{\sigma}}(x)\triangleq\sum_{k=1}^K \frac{\sigma_k^2}{2\theta_k}\left(1-\bbE\left[e^{-2\theta_k Y}\right]e^{-2\theta_kx}\right),
\end{align}
and $\beta^*$ corresponds to the optimal long-term average sum MSE in this case, and is given by the unique solution of
\begin{align} \label{eq:nameless-thm1}
    &p(\beta^*)=\sum_{k=1}^K \frac{\sigma_k^2}{2\theta_k} \left(H(\tau^*_{\text{MAF}})+\frac{K}{\mu(1-\epsilon)}-\frac{1}{2\theta_k}\cdot\frac{\mu}{2\theta_k+\mu}(1-F_k(\tau^*_{\text{MAF}}))\right) \nonumber \\
    &\hspace{3in}-\beta^*\left(H(\tau^*_{\text{MAF}})+\frac{K}{\mu(1-\epsilon)}\right)=0,
\end{align}
in which $H(\cdot)$ and $F_k(\cdot)$ are defined as follows:
\begin{align}
    H(\tau)=\sum_{\rho=K}^{\infty} \binom{\rho-1}{K-1} \epsilon^{\rho-K} (1-\epsilon)^K \left[\tau \gamma(\mu\tau,\rho) -\frac{\rho}{\mu}\gamma(\mu\tau,\rho+1)\right],
\end{align}
\begin{align}
    F_k(\tau)=\sum_{\rho=K}^\infty \binom{\rho\!-\!1}{K\!-\!1} \epsilon^{\rho-K} (1\!-\!\epsilon)^K \left[e^{-2\theta_k \tau} \gamma(\mu \tau, \rho)\!+\!\left(\frac{\mu}{2\theta_k+\mu}\right)^{\rho} (1\!-\!\gamma((2\theta_k\!+\!\mu)\tau, \rho)\right],
\end{align}
where $\gamma(x,y)$ is the normalized incomplete Gamma function: $\gamma(x,y)=\frac{1}{(y-1)!} \int_0^x t^{y-1} e^{-t} dt$. 
\end{theorem}

\begin{Proof}
We follow Dinkelbach's approach \cite{dinkelbach-fractional-prog} to solve the optimization problem in \eqref{sumMSE}. We start by defining an auxiliary function $p(\beta)$, for $\beta \geq 0$, such that:
\begin{align}
    p(\beta)=\min_{w(\cdot)\geq 0} &\quad \sum_{k=1}^K \bbE\left[\int_{D}^{D+\Gamma(K,\epsilon,w)}\texttt{mse}_{\text{MAF}}^{[k]}(t,\tilde{S}_i^{[k]})dt\right]-\beta \bbE\left[\Gamma(K,\epsilon,w)\right]\notag\\
    \text{s.t.} &\quad \bbE\left[(1-\epsilon)w\left(\tilde{Y}\right)+\sum_{k=1}^K Y^{[k]}\right] \geq \frac{K}{f_{\max}}
\end{align}
The optimal solution of our original optimization problem \eqref{sumMSE} is the solution of $p(\beta^*)=0$ of the auxiliary problem above \cite{dinkelbach-fractional-prog}.

The Lagrangian corresponding to the auxiliary problem can be written as:
\begin{align}\label{Lagrangian_sum}
    \mathcal{L}=&\sum_{k=1}^K\bbE\left[\int_{D}^{D+\Gamma(K,\epsilon,w)}\texttt{mse}_{\text{MAF}}^{[k]}(t,\tilde{S}_i^{[k]})dt\right]-\beta \bbE\left[\Gamma(K,\epsilon,w)\right]-\int_0^{\infty} \eta(y)w(y)dy \notag\\
    &-\zeta \left(\bbE\left[(1-\epsilon)w\left(\tilde{Y}\right)+\sum_{k=1}^K Y^{[k]}\right] - \frac{K}{f_{\max}}\right) 
\end{align}
where the last term in the first line of (37), namely, $\int_{0}^{\infty} \eta(y) w(y)$ corresponds to the non-negativity constraint on the waiting function $w(\cdot)$. Now, denoting $\tilde{Y}=\sum_{k=1}^K \sum_{m=1}^{M^{[k]}} Y^{[k]}(m)$, we calculate the expected epoch length as:
\begin{align}
    \bbE\left[\Gamma(K,\epsilon,w)\right]=&\bbE\left[w(\tilde{Y})+\sum_{k=1}^K \tilde{Y}\right]\\
    =&\bbE\left[w(\tilde{Y})\right]+\sum_{k=1}^K \bbE\left[M^{[k]}\right]\bbE\left[Y^{[k]}\right]\label{epoch_wald}\\
    =&\bbE\left[w(\tilde{Y})\right]+\frac{K}{\mu(1-\epsilon)}
\end{align}
where \eqref{epoch_wald} follows from Wald's identity. The expected total MSE in the epoch can be calculated as:
\begin{align}
    &\bbE\left[\int_{D_i}^{D_i+\Gamma(K,\epsilon,w)} \frac{\sigma_k^2}{2\theta_k}\left(1-e^{-2\theta_k\left(t-\tilde{S}_i^{[k]}\right)}\right) dt\right]\notag\\
    &\hspace{.5in}=\frac{\sigma_k^2}{2\theta_k} \left(\bbE\left[\Gamma(K,\epsilon,w)\right]-\frac{1}{2\theta_k} \bbE\left[e^{2\theta_k \tilde{S}_i^{[k]}}\left(e^{-2\theta_k D}-e^{-2\theta_k(D+\Gamma(K,\epsilon,w))}\right)\right]\right)\\
    &\hspace{.5in}=\frac{\sigma_k^2}{2\theta_k} \left(\bbE\left[\Gamma(K,\epsilon,w)\right]-\frac{1}{2\theta_k} \bbE\left[e^{-2\theta_k Y_0}\left(1-e^{-2\theta_k\Gamma(K,\epsilon,w)}\right)\right]\right)\\
    &\hspace{.5in}=\frac{\sigma_k^2}{2\theta_k} \left(\bbE\left[\Gamma(K,\epsilon,w)\right]-\frac{1}{2\theta_k} \bbE\left[e^{-2\theta_k Y}\right]\left(1-\bbE\left[e^{-2\theta_k\Gamma(K,\epsilon,w)}\right)\right]\right).\label{mse_final}
\end{align}

We note the following functional derivatives with respect to $w(\cdot)$ (at realization $\tilde{Y}=z$), 
\begin{align}
    &\frac{\partial}{\partial w(\cdot)} \bbE[w(\tilde{Y})]=f_{\tilde{Y}}(z),\\
    &\frac{\partial}{\partial w(\cdot)} \bbE\left[e^{-2\theta_k\Gamma(K,\epsilon,w)}\right]=\frac{\partial}{\partial w(\cdot)} \bbE\left[e^{-2\theta_k(w(\tilde{Y})+\tilde{Y})}\right]=-2\theta_k e^{-2\theta_k (w(z)+z)}f_{\tilde{Y}}(z),\\
    &\frac{\partial}{\partial w(\cdot)} \int_0^{\infty} \eta(y) w(y)dy=\eta(z).
\end{align}
By substituting in \eqref{Lagrangian_sum} and use the stationarity condition of the Lagrangian with respect to the functional $w(\cdot)$, i.e., $\frac{\partial \mathcal{L}}{\partial w(z)}=0$, we get
\begin{align}
    \sum_{k=1}^K \frac{\sigma_k^2}{2\theta_k}\left(f_{\tilde{Y}}(z)-\bbE\left[e^{-2\theta_k Y} \right]e^{-2\theta_k(w^*(z)+z)} f_{\tilde{Y}}(z)\right)-\beta f_{\tilde{Y}}(z)-\eta(z)-\zeta(1-\epsilon) f_{\tilde{Y}}(z)=0,
\end{align}
which leads to
\begin{align}
  \sum_{k=1}^K \frac{\sigma_k^2}{2\theta_k}\left( 1-\bbE\left[e^{-2\theta_k Y} \right]e^{-2\theta_k(w^*(z)+z)}\right)= \beta+\zeta(1-\epsilon)+\frac{\eta(z)}{f_{\tilde{Y}}(z)}. 
\end{align}

Now define the function $G_{\boldsymbol{\theta},\boldsymbol{\sigma}^2}(\cdot)$ as
\begin{align}
    G_{\boldsymbol{\theta},\boldsymbol{\sigma}^2}(x)=\sum_{k=1}^K\frac{\sigma_k^2}{2\theta_k}\left( 1-\bbE\left[e^{-2\theta_k Y} \right]e^{-2\theta_kx}\right).
\end{align}
Such function is monotonically increasing. Consequently, the equation $G_{\boldsymbol{\theta},\boldsymbol{\sigma}^2}(w^*(z)+z)=\beta+\zeta(1-\epsilon)+\frac{\eta(y)}{f_{\tilde{Y}}(z)}$ has a unique solution, which is given by
\begin{align}
    w^*(z)+z=G_{\boldsymbol{\theta},\boldsymbol{\sigma}^2}^{-1}\left(\beta+\zeta(1-\epsilon)+\frac{\eta(y)}{f_{\tilde{Y}}(z)}\right).
\end{align}
Using the complementary slackness for the constraint $w(\cdot) \geq 0$ \cite{boyd}, we have,
\begin{align}
    w^*(z)=\left[G_{\boldsymbol{\theta},\boldsymbol{\sigma}^2}^{-1}\left(\beta+\zeta(1-\epsilon)\right)-z\right]^+.
\end{align}

This proves the first part of the theorem, which states that the optimal waiting function is in fact a threshold policy with a threshold
\begin{align}
    \tau^*_{\text{MAF}}(K,f_{\max},\epsilon,\boldsymbol{\theta},\boldsymbol{\sigma}^2)=G_{\boldsymbol{\theta},\boldsymbol{\sigma}^2}^{-1}\left(\beta+\zeta(1-\epsilon)\right)
\end{align}

Now, we focus on characterizing the Lagrange multiplier $\zeta$. First, if $f_{\max} \geq \mu$, then $\bbE[w(\tilde{Y})] \geq 0$, which is feasible for any $w(\cdot) \geq 0$. Consequently, in this case the sampling constraint is never active and $\zeta^*=0$, i.e.,
\begin{align}
    \tau^*_{\text{MAF}}(K,f_{\max},\epsilon,\boldsymbol{\theta},\boldsymbol{\sigma}^2)=G_{\boldsymbol{\theta},\boldsymbol{\sigma}^2}^{-1}\left(\beta\right),\quad\text{if $f_{\max} \geq \mu$}.
\end{align}

Next, we consider the other case, i.e., when $f_{\max} < \mu$. In this case, we need to evaluate the expected value of the waiting function $\bbE[w(\tilde{Y})]$ with an arbitrary threshold $\tau$. We denote this expected value by $H(\tau)$:
\begin{align}\label{H_tau}
    H(\tau)= \bbE\left[w(\tilde{Y})\right]=\bbE\left[(\tau-\tilde{Y})^+\right].
\end{align}
To evaluate \eqref{H_tau}, we need to characterize the statistical distribution of $\tilde{Y}$. By the law of total probability, we can write,
\begin{align}
    f_{\tilde{Y}}(z)=\sum_{\rho=K}^\infty f_{\tilde{Y}|\sum_{k=1}^K M^{[k]}}\left(z\left|\sum_{k=1}^K M^{[k]}=\rho\right.\right) \mathbb{P}\left(\sum_{k=1}^K M^{[k]}=\rho\right).
\end{align}
Conditioned on $\sum_{k=1}^K M^{[k]}=\rho$, then $\tilde{Y}=\sum_{k=1}^K \sum_{m=1}^{M^{[k]}} Y^{[k]}(m)$ is a sum of $\rho$ i.i.d. exponential random variables. Hence, $\tilde{Y}|\rho \sim \text{Erlang}(\rho,\mu)$, i.e.,
\begin{align}
    f_{\tilde{Y}|\sum_{k=1}^K M^{[k]}}\left(z|\rho\right)=\frac{\mu^\rho z^{\rho-1} e^{-\mu z}}{(\rho-1)!}, \quad z \geq 0.
\end{align}
Further, the distribution of $\sum_{k=1}^K M^{[k]}$ is given by
\begin{align}
    \mathbb{P}\left(\sum_{k=1}^K M^{[k]}=\rho\right)=\binom{\rho-1}{K-1} \epsilon^{\rho-K} (1-\epsilon)^K, \quad \rho \geq K. 
\end{align}
Substituting the above in \eqref{H_tau} leads to
\begin{align}
    H(\tau)&=\int_{0}^\tau (\tau-z)  f_{\tilde{Y}}(z) dz \\
           &=\sum_{\rho=K}^{\infty} \binom{\rho-1}{K-1} \epsilon^{\rho-K} (1-\epsilon)^K \frac{\mu^\rho}{(\rho-1)!} \left[\tau \int_0^\tau z^{\rho-1} e^{-\mu z}dz -\int_0^\tau z^{\rho} e^{-\mu z}dz\right] \label{DCT}\\
           &=\sum_{\rho=K}^{\infty} \binom{\rho-1}{K-1} \epsilon^{\rho-K} (1-\epsilon)^K \frac{\mu^\rho}{(\rho-1)!} \left[\tau \frac{(\rho-1)!}{\mu^\rho} \gamma(\mu\tau,\rho) -\frac{\rho!}{\mu^{\rho+1} }\gamma(\mu\tau,\rho+1)\right]\\
           &=\sum_{\rho=K}^{\infty} \binom{\rho-1}{K-1} \epsilon^{\rho-K} (1-\epsilon)^K \left[\tau \gamma(\mu\tau,\rho) -\frac{\rho}{\mu}\gamma(\mu\tau,\rho+1)\right],
\end{align}
where interchanging sum and integral in \eqref{DCT} follows from the dominated convergence theorem and $\gamma(x,y)$ is the normalized incomplete Gamma function:
\begin{align}
    \gamma(x,y)=\frac{1}{(y-1)!} \int_0^x t^{y-1} e^{-t} dt
\end{align}

Now observe that the sampling constraint in \eqref{sampling_constraint} can be written as:
\begin{align}
    (1-\epsilon)H(\tau) \geq \frac{K}{f_{\max}}-\frac{K}{\mu}.
\end{align}
Thus, if $(1-\epsilon)H\left(G_{\boldsymbol{\theta},\boldsymbol{\sigma}^2}^{-1}\left(\beta\right)\right) > \frac{K}{f_{\max}}-\frac{K}{\mu}$, then $\zeta^*=0$ and the threshold for the unconstrained problem (without the sampling constraint) is optimal.

Otherwise, the constraint is satisfied with equality, i.e.,$(1-\epsilon)H(\tau) = \frac{K}{f_{\max}}-\frac{K}{\mu}$, and hence,
\begin{align}
    \tau^* =\frac{1}{(1-\epsilon)} H^{-1}\left( \frac{K}{f_{\max}}-\frac{K}{\mu}\right).
\end{align}

All this put together implies that
\begin{align}
    \tau^*_{\text{MAF}}(K,f_{\max},\epsilon,\boldsymbol{\theta},\boldsymbol{\sigma}^2)=\max\left\{G_{\boldsymbol{\theta},\boldsymbol{\sigma}^2}^{-1}\left(\beta\right),\frac{1}{(1-\epsilon)} H^{-1}\left( \frac{K}{f_{\max}}-\frac{K}{\mu}\right)\right\}.
\end{align}

Finally, we solve for the optimal long-term MMSE, $\beta^*$. Denote $\bbE[e^{-2\theta_k \Gamma(K,\epsilon, w)}]$ by $F(\tau)$. By observing that
\begin{align}
    F_k(\tau)&=\bbE[e^{-2\theta_k \Gamma(K,\epsilon, w)}]\\
    &=\int_0^\tau e^{-2\theta_k \tau} f_{\tilde{Y}}(z)dz+\int_\tau^\infty e^{-2\theta_k z} f_{\tilde{Y}}(z)dz\\
    &=\sum_{\rho=K}^\infty \binom{\rho-1}{K-1} \epsilon^{\rho-K} (1-\epsilon)^K \frac{\mu^\rho}{(\rho-1)!} \left[e^{-2\theta_k \tau}\!\!\! \int_0^\tau \!\!\!z^{\rho-1} e^{-\mu z} dz\!+\!\!\int_\tau^\infty \!\!\!z^{\rho-1} e^{-(2\theta_k+\mu)z} dz\right]\\
    &=\sum_{\rho=K}^\infty \binom{\rho-1}{K-1} \epsilon^{\rho-K} (1-\epsilon)^K \left[e^{-2\theta_k \tau} \gamma(\mu \tau, \rho)\!+\!\left(\frac{\mu}{2\theta_k+\mu}\right)^{\rho} (1\!-\!\gamma((2\theta_k\!+\!\mu)\tau, \rho)\right],
\end{align}
$\beta^*$ can be obtained by solving the auxiliary problem $p(\beta^*)=0$, i.e., solving \eqref{eq:nameless-thm1}.

This concludes the proof.
\end{Proof}

\begin{remark} \label{remark_thm1}
 We observe that the optimal threshold $\tau^*$ in \eqref{threshold} is increasing with the erasure probability $\epsilon$. This is indeed the case since both $G^{-1}_{\boldsymbol{\theta},\boldsymbol{\sigma}}(\cdot)$ and $H^{-1}(\cdot)$ are increasing functions, and both of their arguments are increasing with $\epsilon$ (the optimal long-term average sum MSE $\beta^*$, for instance, can only increase with higher erasure rates). This can be attributed to the fact that for higher erasure probabilities, the average samples' inter-delivery time increases. This, in turn, causes the transmitter to wait more before sending a new sample to make sure that the new sample is sufficiently \textbf{different} from the previously delivered sample; this new sample will be used to estimate the signal over a large period of time because of the large expected inter-delivery time caused by the high erasure probability. In this case, the transmitter will be blocked from generating new samples waiting for the previous sample to be delivered. On the other hand, and for smaller erasure probabilities, we can see that the waiting time decreases. Again, this can be attributed to the fact that for smaller erasure probabilities, the average samples' delivery time decreases. This means that when the transmitter has a chance to transmit a sample, it would be better to do so as this sample will be delivered quickly and it will not block the transmitter from sending new samples.

 Note that the fact that second term in the $\max$ function, $H^{-1}$, is increasing with $\epsilon$ is consistent with the observation in Remark~\ref{rmrk_smpl}; the higher the erasure rate, the larger the waiting time should be to compensate for the high rate of back-to-back transmissions under MAF scheduling that takes a toll on the sampling frequency in this case.
 \end{remark}

\section{Optimal Waiting Threshold and Minimum Sum MSE Characterization without Erasure Feedback}
In this section, we consider the setting in which the receiver does not provide erasure status feedback to the transmitter. As in Section~\ref{sec: solution presence feedback}, we focus on the sum MSE penalty together with stationary deterministic waiting policies. 

With a slight abuse of notation, let $\tilde{Y}_i(m)=\sum_{k=1}^K Y_i^{[k]}(m)$ be the sum of the service times during the $m$th transmission round. To develop a stationary deterministic waiting policy, we cannot rely on the starting MSE of the epoch (or the starting AoI) since the transmitter does not know of its exact starting time due to the absence of the erasure feedback. Alternatively, we focus our attention on a waiting time that is a function $w(\cdot)$ of the sum of service times of all processes in the preceding {\it transmission round}, as opposed to the preceding epoch, i.e., we set
\begin{align}
    W_i(m)=w\left(\tilde{Y}_{i}(m-1)\right)=w\left(\sum_{k=1}^K Y_i^{[k]}(m-1)\right),
\end{align}
with $\tilde{Y}_i(0)\triangleq Y_{i-1}(M_{i-1})$ by definition. This policy is realizable in the absence of erasure feedback since the start and end of transmission rounds are known at the transmitter side. Note that this policy is a stationary deterministic waiting policy since $\tilde{Y}_i(m)$'s are i.i.d across all transmission rounds.

Similar to Section~\ref{sec: solution presence feedback}, problem \eqref{opt_gen_RR_onewait} in Lemma~\ref{lemma: reformulation absence of feedback} can be written in the following functional form:
\begin{align}\label{sumMSE_absence}
    \min_{w(\cdot)\geq 0} &\quad \frac{ \sum_{k=1}^K \bbE\left[\int_{D^{[k]}}^{D^{[k]}+\Gamma} \texttt{mse}_{\text{RR}}^{[k]}\left(t,\tilde{S}^{[k]}\right)dt\right]}{\mathbb{E}\left[\Gamma \right]} \notag\\
    \text{s.t.~} &\quad \bbE\left[w\left(\tilde{Y}\right)\right] \geq \frac{K}{f_{\max}}-\frac{K}{\mu}.
\end{align}

Theorem~\ref{thm2} below provides the solution of problem \eqref{sumMSE_absence}. 


\begin{theorem}\label{thm2}
The optimal waiting policy $w^*(\cdot)$ that solves problem (\ref{sumMSE_absence}) is given by the \emph{threshold} policy
\begin{align}\label{waiting-policy-opt}
    w^*(z)=\left[\tau_{\text{RR}}^*(K,f_{\max},\epsilon,\boldsymbol{\theta},\boldsymbol{\sigma})-z\right]^+,
\end{align}
where the optimal threshold $\tau_{\text{RR}}^*(K,f_{\max},\epsilon,\boldsymbol{\theta},\boldsymbol{\sigma})$ is given by
\begin{align}\label{threshold_NoFB}
    \tau_{\text{RR}}^*(K,f_{\max},\epsilon,\boldsymbol{\theta},\boldsymbol{\sigma}^2)=\max\left\{\tilde{G}_{\boldsymbol{\theta},\boldsymbol{\sigma}^2,\epsilon}^{-1}(\beta), \tilde{H}^{-1}\left( \frac{K}{f_{\max}}-\frac{K}{\mu}\right)\right\}
\end{align}
in which
\begin{align}
\tilde{G}_{\boldsymbol{\theta},\boldsymbol{\sigma}^2,\epsilon}(x)\triangleq\sum_{k=1}^K \frac{\sigma_k^2}{2\theta_k}\left(1-\bbE\left[e^{-2\theta_k Y} \right]\frac{(1-\epsilon)^2 e^{-2\theta_k x}}{\left(1-\epsilon L_k(x) \right)^2}\right),
\end{align}
with $L_k(\cdot)$ being
\begin{align}
    L_k(x)\triangleq e^{-2\theta_k x} \gamma(\mu x,K)+\left(\frac{\mu}{\mu+2\theta_k}\right)^K (1-\gamma\left((2\theta_k+\mu) x, K\right),
\end{align}
and $\beta^*$ corresponds to the optimal long-term average sum MSE in this case, and is given by the unique solution of
\begin{align}
    &\tilde{p}(\beta)=\sum_{k=1}^K \frac{\sigma_k^2}{2\theta_k} \left(\frac{\tilde{H}(\tau^*_{\text{RR}})+\frac{K}{\mu}}{1-\epsilon}-\frac{1}{2\theta_k} \cdot \frac{\mu}{\mu+2\theta_k}\cdot\left(1-\tilde{F}_k(\tau^*_{\text{RR}})\right)\right)-\beta\left(\frac{\tilde{H}(\tau^*_{\text{RR}})+\frac{K}{\mu}}{1-\epsilon}\right)=0,
\end{align}
in which $\tilde{H}(\cdot)$ and $\tilde{F}_k(\cdot)$ are defined as follows:
\begin{align}\label{H_fn}
    \tilde{H}(\tau)=\tau \gamma \left(\mu \tau, K\right) -\frac{K}{\mu} \gamma\left(\mu \tau,K+1\right),
\end{align}
\begin{align}\label{F_fn_NoFB}
    \tilde{F}_k(\tau)=\frac{(1-\epsilon)L_k(\tau)}{1-\epsilon L_k(\tau)}
\end{align}
where $\gamma(x,y)$ is the normalized incomplete Gamma function.     
\end{theorem}

\begin{Proof}
Similar to the proof of Theorem~\ref{thm1}, we follow the Dinkelbach's approach to solve the optimization problem in \eqref{sumMSE_absence}. We start by defining the auxiliary function $\tilde{p}(\beta)$ for $\beta \geq 0$, such that:
\begin{align}
    \tilde{p}(\beta)=\min_{w(\cdot)\geq 0} &\quad \sum_{k=1}^K \bbE\left[\int_{D}^{D+\Gamma(K,\epsilon,w)}\texttt{mse}_{\text{RR}}^{[k]}(t,\tilde{S}_i^{[k]})dt\right]-\beta \bbE\left[\Gamma(K,\epsilon,w)\right]\notag\\
    \text{s.t.} &\quad \bbE\left[w\left(\tilde{Y}\right)+\sum_{k=1}^K Y^{[k]}\right] \geq \frac{K}{f_{\max}}
\end{align}
Hence, the Lagrangian function corresponding the auxiliary problem above is given by
\begin{align}\label{Lagrangian_sum_absence}\mathcal{L}=&\sum_{k=1}^K\bbE\left[\int_{D}^{D+\Gamma(K,\epsilon,w)}\texttt{mse}_{\text{RR}}^{[k]}(t,\tilde{S}_i^{[k]})dt\right]-\beta \bbE\left[\Gamma(K,\epsilon,w)\right]-\int_0^{\infty} \eta(y)w(y)dy \notag\\
    &-\zeta \left(\bbE\left[w\left(\tilde{Y}\right)+\sum_{k=1}^K Y^{[k]}\right] - \frac{K}{f_{\max}}\right). 
\end{align}

Without loss of generality, we focus on the epoch of the $K$th process. In this case, we can write $\Gamma_i^{[K]}$ as
\begin{align}
        \Gamma_i^{[K]}=w\left(\tilde{Y}_{i-1}(M_{i-1}^{[K]})\right)+\sum_{m=1}^{M_i^{[K]}-1} \tilde{Y}_i(m)+w\left(\tilde{Y}_{i}(m)\right)+\tilde{Y}_i\left(M_i^{[K]}\right),
\end{align}
where $M_i^{[K]} \geq 2$. Note that $\Gamma_i^{[1]} \sim \cdots  \sim \Gamma_i^{[K]} \sim \Gamma(K,w,\epsilon)$. Moreover, due to the stationarity of the waiting policy, we can drop the indices $i$, and $K$, and re-define $\tilde{Y}_{i-1}^{[K]}(M_{i-1}^{K})=\tilde{\underline{Y}}$, and $\tilde{Y}_{i}^{[K]}(M_{i}^{K})=\tilde{\bar{Y}}$. Hence, we have the expected epoch length given by
    \begin{align}
        \mathbb{E}\left[\Gamma(K,\epsilon,w)\right]&=\mathbb{E}\left[\Gamma_i^{[K]}\right]\\
                          &=\mathbb{E}[w(\tilde{\underline{Y}})]+\mathbb{E}[M-1]\cdot \mathbb{E}[\tilde{Y}+w(\tilde{Y})]+\mathbb{E}[\tilde{\bar{Y}}]\label{epoch_Wald_absence}\\
                          &=\mathbb{E}[M]\cdot \left(\mathbb{E}[\tilde{Y}]+\mathbb{E}[w(\tilde{Y})]\right)\label{epoch_service}\\
                          &=\frac{\frac{K}{\mu}+\mathbb{E}[w(\tilde{Y})]}{1-\epsilon} \label{epoch_M}
    \end{align}
where \eqref{epoch_Wald_absence} follows from Wald's equation, \eqref{epoch_service} follows from the fact that the aggregate service times $\tilde{\underline{Y}} \sim \tilde{Y} \sim \tilde{\bar{Y}} \sim \text{Erlang}(K,\mu)$ as the individual service times are i.i.d. $\sim \text{exp}(\mu)$, and \eqref{epoch_M} follows from the fact that the number of transmission rounds $M \sim \text{geometric}(1-\epsilon)$. From \eqref{mse_final}, we have the expected MSE in the epoch given by
\begin{align}
    \bbE&\left[\int_{D_i}^{D_i+\Gamma(K,\epsilon,w)} \frac{\sigma_k^2}{2\theta_k}\left(1-e^{-2\theta_k\left(t-\tilde{S}_i^{[k]}\right)}\right) dt\right]\notag\\
   & \qquad\qquad\qquad\qquad=\frac{\sigma_k^2}{2\theta_k} \left(\bbE\left[\Gamma(K,\epsilon,w)\right]-\frac{1}{2\theta_k} \bbE\left[e^{-2\theta_k Y}\right]\left(1-\bbE\left[e^{-2\theta_k\Gamma(K,\epsilon,w)}\right)\right]\right).
\end{align}

Due to the stationary policy $w(\cdot)$, and the fact that all service times are i.i.d., we can evaluate the expectation $\mathbb{E}\left[ e^{-2\theta_k \Gamma(K,\epsilon,w)}\right]$ as follows:
\begin{align}
    \mathbb{E}\left[ e^{-2\theta_k \Gamma(K,\epsilon,w)}\right]=& \mathbb{E}\left[ \mathbb{E}\left[e^{-2\theta_k \sum_{m=0}^{M-1} w(\tilde{Y}(m))+\tilde{Y}(m)}|M\right]\right]\label{expectation_epoch1}\\
    =&\sum_{m=1}^\infty \left(\mathbb{E}\left[e^{-2\theta_k (w(\tilde{Y})+\tilde{Y})}\right]\right)^{m} \epsilon^{m-1}(1-\epsilon)\label{expectation_epoch2}\\
    =&\frac{(1-\epsilon) \mathbb{E}\left[e^{-2\theta_k (w(\tilde{Y})+\tilde{Y})}\right]}{1-\epsilon \mathbb{E}\left[e^{-2\theta_k (w(\tilde{Y})+\tilde{Y})}\right]}
\end{align}
where \eqref{expectation_epoch1} follows from iterated expectation over the number of transmission rounds $M$ with $\tilde{Y}(0)=\tilde{\underline{Y}} \sim \tilde{\bar{Y}} $. This leads to the following functional derivative of the expectation $ \mathbb{E}\left[ e^{-2\theta_k \Gamma(K,\epsilon,w)}\right]$ with respect to $w(\cdot)$ (at realization $\tilde{Y}=z$),
\begin{align}
    \frac{\partial}{\partial w(\cdot)}  \mathbb{E}\left[ e^{-2\theta_k \Gamma(K,\epsilon,w)}\right]=\frac{-2\theta_k(1-\epsilon) e^{-2\theta_k (w(z)+z)}f_{\tilde{Y}}(z)}{\left(1-\epsilon \mathbb{E}\left[e^{-2\theta_k (w(\tilde{Y})+\tilde{Y})}\right] \right)^2}.
\end{align}

Applying the stationarity condition of the Lagrangian with respect to the functional $w(\cdot)$, $\frac{\partial \mathcal{L}}{\partial w(z)}=0$, we get the following optimality condition:
\begin{align}
   \sum_{k=1}^K \frac{\sigma_k^2}{2\theta_k}\left(1\!-\!\bbE\left[e^{-2\theta_k Y} \right]\frac{(1-\epsilon)^2 e^{-2\theta_k (w(z)+z)}}{\left(1-\epsilon \mathbb{E}\left[e^{-2\theta_k (w(\tilde{Y})\!+\!\tilde{Y})}\right] \right)^2}\right)=\beta\!+\!\frac{(1-\epsilon)\eta(z)}{f_{\tilde{Y}}(z)}\!+\!(1-\epsilon)\zeta.
\end{align}

Define the function $\tilde{G}_{\boldsymbol{\theta},\boldsymbol{\sigma}^2,\epsilon,w}(\cdot)$ as
\begin{align}
    \tilde{G}_{\boldsymbol{\theta},\boldsymbol{\sigma}^2,\epsilon,w}(x)=\sum_{k=1}^K \frac{\sigma_k^2}{2\theta_k}\left(1\!-\!\bbE\left[e^{-2\theta_k Y} \right]\frac{(1-\epsilon)^2 e^{-2\theta_k x}}{\left(1-\epsilon \mathbb{E}\left[e^{-2\theta_k (w(\tilde{Y})\!+\!\tilde{Y})}\right] \right)^2}\right)
\end{align}
Observe that the function $G_{\boldsymbol{\theta},\boldsymbol{\sigma}^2,\epsilon,w}$ is a monotonically increasing function in $x$ (given that $\mathbb{E}\left[e^{-2\theta_k (w(\tilde{Y})\!+\!\tilde{Y})}\right]$ for any specific choice of $w(\cdot)$ is fixed, irrespective of the realizations). Thus, using the complementary slackness condition, the optimal waiting function is indeed a threshold waiting policy in the form of \footnote{The optimal threshold seems to be \emph{self-dependent} through the expectation term $\mathbb{E}\left[e^{-2\theta_k (w(\tilde{Y})\!+\!\tilde{Y})}\right]$. This does not affect the claim that the optimal waiting policy is a threshold policy, as the function $\tilde{G}_{\boldsymbol{\theta},\boldsymbol{\sigma}^2,\epsilon,w}(\cdot)$ is a monotone function for any choice of $w(\cdot)$, which further implies the existence of a unique solution. Nevertheless, finding this threshold numerically would require iterating back and forth between the expectation term $L_k(\tau)$ (see \eqref{F2}) and the inverse function $\tilde{G}_{\boldsymbol{\theta},\boldsymbol{\sigma}^2,\epsilon,w}^{-1}(\cdot)$ as we show later on in the proof.}
    \begin{align}
    w^*(z)=\left[\tilde{G}_{\boldsymbol{\theta},\boldsymbol{\sigma}^2,\epsilon,w}^{-1}\left(\beta\!+\!(1-\epsilon)\zeta\right)-z\right]^+.
\end{align}

Now that the threshold behavior has been established, we note that the threshold policy maintains the monotonicity behavior of $\tilde{G}(\cdot)$ since $\mathbb{E}\left[e^{-2\theta_k (w(\tilde{Y})\!+\!\tilde{Y})}\right]$ would be a monotone decreasing function in the threshold value.

Considering a threshold waiting policy of the form $w(\tilde{y})=\left[\tau-\tilde{y}\right]^+$, and since $\tilde{Y} \sim \text{Erlang}(K,\mu)$, we can evaluate the expectation in $\mathbb{E}\left[e^{-2\theta_k (w(\tilde{Y}+\tilde{Y}))}\right]$, denoted by $L_k(\tau)$, as follows:
\begin{align}\label{F2}
    L_k(\tau)\triangleq\mathbb{E}\left[e^{-2\theta_k (w(\tilde{Y}+\tilde{Y}))}\right]&=\int_0^{\tau} e^{-2\theta_k \tau} \frac{\mu^K \tilde{y}^{K-1} e^{-\mu \tilde{y}}}{(K-1)!} d\tilde{y}+\int_{\tau}^\infty e^{-2\theta_k \tilde{y}} \frac{\mu^K \tilde{y}^{K-1} e^{-\mu \tilde{y}}}{(K-1)!} d\tilde{y}\\
    &= e^{-2\theta_k \tau} \gamma(\mu \tau,K)+\frac{\mu^K}{(K-1)!}\int_{\tau}^\infty \tilde{y}^{K-1} e^{-(2\theta_k+\mu)} d\tilde{y}\\
    &=e^{-2\theta_k \tau} \gamma(\mu \tau,K)+\left(\frac{\mu}{\mu+2\theta_k}\right)^K (1-\gamma\left((2\theta_k+\mu) \tau), K\right).
\end{align}
In this case, the function $\tilde{G}_{\boldsymbol{\theta},\boldsymbol{\sigma}^2,\epsilon}(\cdot)$ (we dropped the $w(\cdot)$ dependence as we deal with a threshold structure) can be re-written as:
    \begin{align}
\tilde{G}_{\boldsymbol{\theta},\boldsymbol{\sigma}^2,\epsilon}(x)=\sum_{k=1}^K \frac{\sigma_k^2}{2\theta_k}\left(1\!-\!\bbE\left[e^{-2\theta_k Y} \right]\frac{(1-\epsilon)^2 e^{-2\theta_k x}}{\left(1-\epsilon L_k(x) \right)^2}\right).
\end{align}
We also evaluate the expected waiting time function, $\tilde{H}(\tau)$, as
\begin{align}\label{H_absence}
    \tilde{H}(\tau)\triangleq\mathbb{E}[w(\tilde{Y})]&=\mathbb{E}\left[\left[\tau-\tilde{Y}\right]^+\right]\\
    &= \int_{0}^{\tau} (\tau-\tilde{y}) f(\tilde{y}) d\tilde{y}\\
    &=\tau \int_{0}^{\tau}  \frac{\mu^K \tilde{y}^{K-1} e^{-\mu \tilde{y}}}{(K-1)!} d\tilde{y}-\int_{0}^\tau \tilde{y}  \frac{\mu^K \tilde{y}^{K-1} e^{-\mu \tilde{y}}}{(K-1)!} d\tilde{y}\\
    &= \tau \gamma \left(\mu \tau, K\right) -\frac{K}{\mu} \gamma\left(\mu \tau,K+1\right).
\end{align}
Finally, for compactness, we define 
\begin{align}
    \tilde{F}_k(\tau)\triangleq \bbE\left[e^{-2\theta_k \Gamma(K,\epsilon, w)}\right]=\frac{(1-\epsilon) L_k(\tau)}{1-\epsilon L_k(\tau)}.
\end{align}

Plugging all these quantities into the long-term average MSE penalty, we get
\begin{align}
    \sum_{k=1}^K \overline{\texttt{mse}_{\text{RR}}^{[k]}}&=\frac{ \sum_{k=1}^K \bbE\left[\int_{D^{[k]}}^{D^{[k]}+\Gamma} \texttt{mse}_{\text{RR}}^{[k]}\left(t,\tilde{S}^{[k]}\right)dt\right]}{\mathbb{E}\left[\Gamma \right]}\\
    &=\sum_{k=1}^K \frac{\sigma_k^2}{2\theta_k} \left(1-\frac{1}{2\theta_k} \cdot\frac{\bbE\left[e^{-2\theta_k Y}\right]}{\bbE[\Gamma(K,\epsilon,w)]}\left(1-\bbE\left[e^{-2\theta_k \Gamma(K,\epsilon,w)}\right]\right)\right)\\
    &=\sum_{k=1}^K \frac{\sigma_k^2}{2\theta_k} \left(1-\frac{1}{2\theta_k} \cdot \frac{\mu}{\mu+2\theta_k}\cdot\frac{1-\epsilon}{\tilde{H}(\tau)+\frac{K}{\mu}}\cdot\left(1-\tilde{F}_k(\tau)\right)\right)
\end{align}
Hence, the optimal solution the auxiliary problem, $\tilde{p}(\beta^*)=0$, is given by solving
\begin{align}
    \sum_{k=1}^K \frac{\sigma_k^2}{2\theta_k} \left(\frac{\tilde{H}(\tau)+\frac{K}{\mu}}{1-\epsilon}-\frac{1}{2\theta_k} \cdot \frac{\mu}{\mu+2\theta_k}\cdot\left(1-\tilde{F}_k(\tau)\right)\right)-\beta^*\left(\frac{\tilde{H}(\tau)+\frac{K}{\mu}}{1-\epsilon}\right)=0
\end{align}
as stated in the theorem.

Focusing on the sampling frequency constraint, one can rewrite it as
\begin{align}
    \tilde{H}(\tau) \geq \frac{K}{f_{\max}}-\frac{K}{\mu}.
\end{align}
Hence, similar to the proof of Theorem~\ref{thm1}, if the sampling constraint is non-binding, i.e., if $\tilde{H}(\tilde{G}_{\boldsymbol{\theta},\boldsymbol{\sigma}^2,\epsilon}^{-1}(\beta))>\frac{K}{f_{\max}}-\frac{K}{\mu}$, the unconstrained solution of the threshold is optimal. Otherwise, we satisfy the constraint with equality, i.e., we set
\begin{align}
    \tau_{\text{RR}}^*=\tilde{H}^{-1}\left(\frac{K}{f_{\max}}-\frac{K}{\mu}\right).
\end{align}
Combining the above two cases gives \eqref{threshold_NoFB}, and completes the proof.
\end{Proof}

\begin{remark} \label{remark_thm2}
    Different from Theorem~\ref{thm1}, the function $\tilde{G}_{\boldsymbol{\sigma},\boldsymbol{\theta},\epsilon}$ is not necessarily increasing in $\epsilon$ as the numerator and denominator are both monotonically decreasing functions in $\epsilon$. Consequently, the optimal threshold in the absence of an erasure status feedback cannot be always increasing in $\epsilon$ as in Theorem~\ref{thm1}. In fact, our numerical evaluation of the optimal threshold in the absence erasure status feedback shows that the optimal threshold is indeed decreasing as the erasure probability $\epsilon$ increases. This is intuitive as the transmitter, in this case, has no knowledge about the current age of the processes' at the receiver other than it is on average increasing in $\epsilon$. Consequently, waiting less is more conservative and leads to a reduction in the overall long-term average of the sum MSE.
\end{remark}

In Algorithm~\ref{alg:Thm2}, we illustrate how to evaluate the optimal policy in Theorem~\ref{thm2} using a \emph{nested bisection} method. Specifically, we run an outer bisection search over the sum MSE value, $\beta$, and an inner bisection search over the threshold value, $\tau$. For the inner bisection, we efficiently evaluate the inverse function $\tilde{G}^{-1}(\cdot)$ or $\tilde{H}^{-1}(\cdot)$, while the outer bisection solves for the unique solution of the auxiliary problem $\tilde{p}(\beta)=0$.  

\begin{remark}
    Algorithm~\ref{alg:Thm2} can be used to evaluate Theorem~\ref{thm1} as well after replacing $\tilde{G}(\cdot)$ by $G(\cdot)$, $\tilde{H}(\cdot)$ by $H(\cdot)$, $\tilde{F}(\cdot)$ by $F(\cdot)$, $\tilde{p}(\cdot)$ by $p(\cdot)$, and the condition for satisfying the sampling frequency constraint from  $\tilde{H}(\tau_0)<\frac{K}{f_{\max}}-\frac{K}{\mu}$ to $H(\tau_0)<\frac{1}{1-\epsilon}\left(\frac{K}{f_{\max}}-\frac{K}{\mu}\right)$.
\end{remark}

\begin{algorithm}
\caption{Nested Bisection Search for Evaluating the Optimal Policy in Theorem~2}\label{alg:Thm2}
\begin{algorithmic}
\Require  $K$, $\boldsymbol{\sigma}$, $\boldsymbol{\theta}$ $f_{\max}$, $\mu$, $\epsilon$, $\delta$, $\tau_{\max}$
\State $\beta_\ell \gets 0$, and $\beta_h \gets \sum_{k=1}^K \frac{\sigma_k^2}{2\theta_k}$ 
\While{$\beta_h-\beta_\ell > \delta$}\Comment{Outer bisection over the sum MSE value} 
\State $\beta \gets \frac{\beta_h+\beta_\ell}{2}$
\State $\tau_\ell \gets 0$, and $\tau_h \gets \tau_{\max}$
\While{$\tau_h-\tau_\ell > \delta$}\Comment{Inner bisection over the unconstrained threshold} 
\State $\tau \gets \frac{\tau_h+\tau_\ell}{2}$
\State $\mathcal{O}_\tau \gets \tilde{G}(\tau)-\beta$
\If{$\mathcal{O}_\tau>0$}
    \State $\tau_h \gets \tau$  
\Else
    \State $\tau_\ell \gets \tau$
\EndIf
\EndWhile

\Return $\tau_0 \gets \tau$\Comment{Unconstrained solution of the threshold} 
\If{$\tilde{H}(\tau_0)<\frac{K}{f_{\max}}-\frac{K}{\mu}$} \Comment{Checking the sampling frequency constraint}
\State $\tau_\ell \gets \tau_0$, and $\tau_h \gets \tau_{\max}$
\While{$\tau_h-\tau_\ell > \delta$}\Comment{Inner bisection over the constrained threshold}
\State $\tau \gets \frac{\tau_h+\tau_\ell}{2}$
\State $\mathcal{O}_\tau \gets \tilde{H}(\tau)-\left(\frac{K}{f_{\max}}-\frac{K}{\mu}\right)$
\If{$\mathcal{O}_\tau>0$}
    \State $\tau_h \gets \tau$ 
\Else
    \State $\tau_\ell \gets \tau$
\EndIf
\EndWhile
\Else  
\State $\tau \gets \tau_0$
\EndIf

\Return $\tau^*=\tau$

 $\mathcal{O}_\beta \gets \tilde{p}(\beta)=\sum_{k=1}^K \frac{\sigma_k^2}{2\theta_k} \left(\frac{\tilde{H}(\tau)+\frac{K}{\mu}}{1-\epsilon}-\frac{1}{2\theta_k} \cdot \frac{\mu}{\mu+2\theta_k}\cdot\left(1-\tilde{F}_k(\tau)\right)\right)-\beta\left(\frac{\tilde{H}(\tau)+\frac{K}{\mu}}{1-\epsilon}\right)$
\If{$\mathcal{O}_\beta>0$}
    \State $\beta_h \gets \beta$ 
\Else
    \State $\beta_\ell \gets \beta$
\EndIf

\EndWhile

\Return $\beta^*=\beta$
\end{algorithmic}
\end{algorithm}


\section{Numerical Results: Optimal Threshold Behavior}



In this section, we present our numerical results concerning Theorem~\ref{thm1} and Theorem~\ref{thm2}. We study the effects of the erasure probability, the number of users, and the speed of the processes on the optimal waiting threshold and the MMSE with and without erasure status feedback under MAF or RR scheduling, respectively.

\subsection{Effect of Erasure Probability}
 \begin{figure}[t]
     \centering
     \begin{subfigure}[b]{0.48\textwidth}
         \centering
         \includegraphics[width=\textwidth]{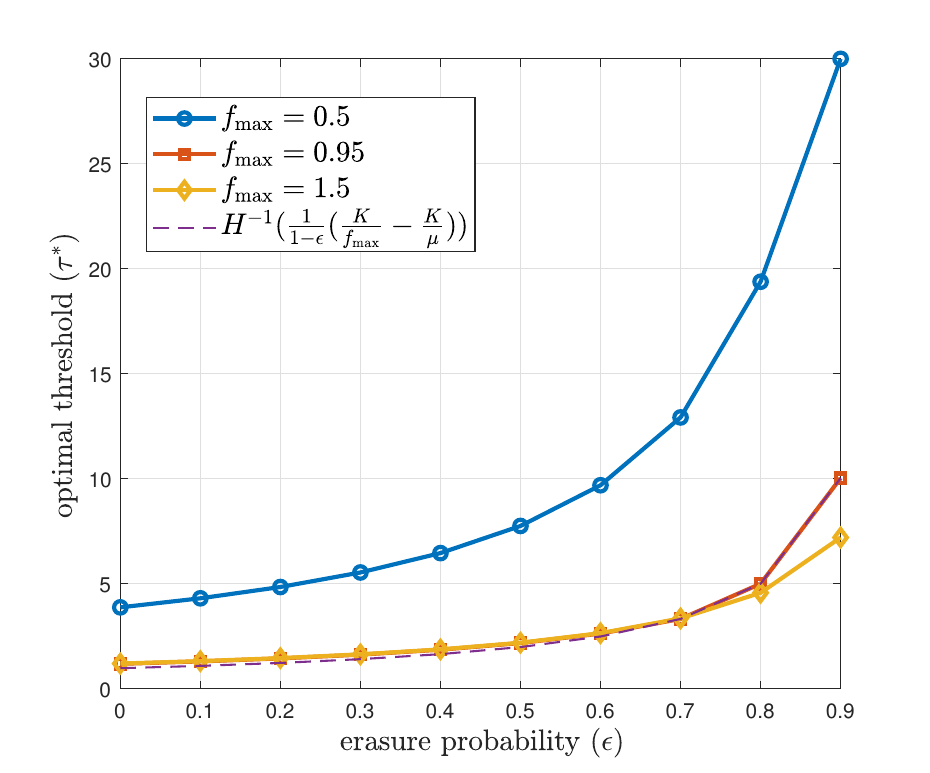}
         \caption{erasure status feedback is present and MAF scheduling.}
         \label{fig:threshold_epsilon}
     \end{subfigure}
     \hfill
     \begin{subfigure}[b]{0.48\textwidth}
         \centering
         \includegraphics[width=\textwidth]{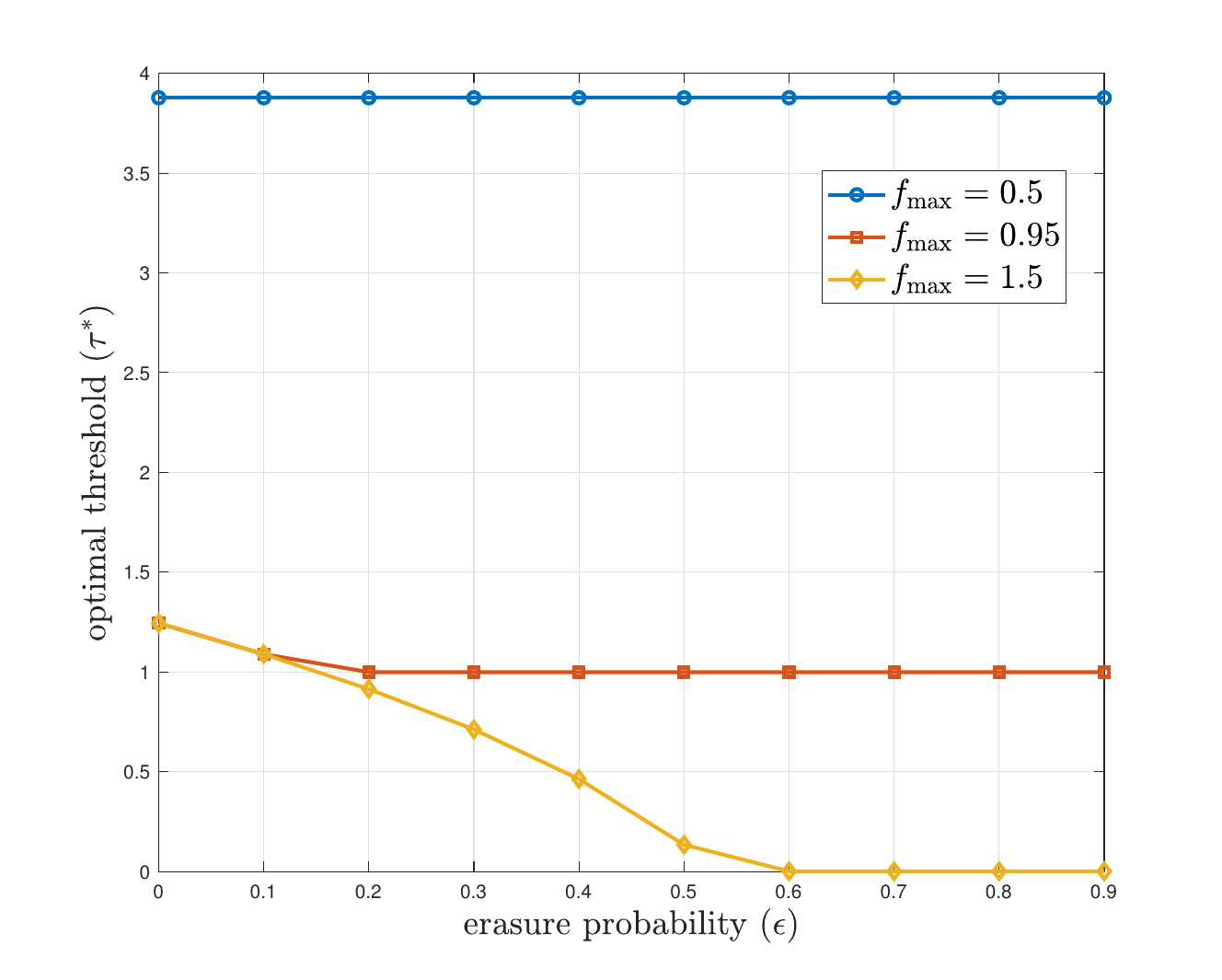}
         \caption{erasure status feedback is absent and RR scheduling.}
         \label{fig:threshold_epsilon_NoFB}
     \end{subfigure}
     \caption{The optimal threshold ($\tau^*_{\text{MAF}}$ (right) and $\tau^*_{\text{RR}}$ (left)) versus the erasure probability ($\epsilon$) for different sampling frequency constraints ($f_{\max}$).}
\end{figure}
We study a 2-process system with $\boldsymbol{\theta}=[0.1 \quad 0.5]$, and $\boldsymbol{\sigma}=[1 \quad 2]$. The exponential service rate is $\mu=1$. In case of the presence/absence of the erasure status feedback, we study how the optimal threshold $\tau^*_{\text{MAF}}/\tau^*_{\text{RR}}$ behaves versus the erasure probability $\epsilon$ for $f_{\max}=0.5$, $0.95$, and $1.5$.

First, we consider the case with erasure status feedback in Fig.~\ref{fig:threshold_epsilon}. Our results show that for all sampling frequency constraints, the optimal threshold {\it increases} as the erasure probability increases. This is due to the fact that the functions $G_{\boldsymbol{\theta},\boldsymbol{\sigma}}^{-1}(\cdot)$ and $H^{-1}(\cdot)$ are increasing functions in their argument, which are, in turn, increasing functions of $\epsilon$ (see Remark~\ref{remark_thm1}). Nevertheless, we have three different cases. First, when $f_{\max}=0.5$, the sampling frequency constraint is binding even at $\epsilon=0$. Hence, the optimal threshold is given by
\begin{align}
H^{-1}\left(\frac{1}{1-\epsilon}\left[\frac{K}{f_{\max}}-\frac{K}{\mu}\right]\right)=H^{-1}\left(\frac{2}{1-\epsilon}\right).    
\end{align}
We see that the optimal threshold is higher than the other two cases and has a much steeper curve versus $\epsilon$. On the other hand, when $f_{\max}=1.5$, the sampling frequency constraint is inactive since $f_{\max}>\mu$, and the optimal threshold is given by $G^{-1}_{\boldsymbol{\theta},\boldsymbol{\sigma}}(\beta^*)$ for all $\epsilon$. Finally, for the case when $f_{\max}=0.95$, we observe an interesting behavior: when $\epsilon<\epsilon^*=0.7$, the threshold corresponding to $G^{-1}_{\boldsymbol{\theta},\boldsymbol{\sigma}}(\beta^*)$ is (slightly) higher than the threshold corresponding to  $H^{-1}\left(\frac{1}{1-\epsilon}\left[\frac{K}{f_{\max}}-\frac{K}{\mu}\right]\right)$ (which is shown as a dotted curve in Fig.~\ref{fig:threshold_epsilon}); while for $\epsilon>\epsilon^*=0.7$, the sampling frequency constraint becomes binding and therefore, the optimal threshold is characterized by $H^{-1}(\cdot)$ and becomes steeper.

\begin{figure}[t]
\center
\includegraphics[width=0.8\textwidth]{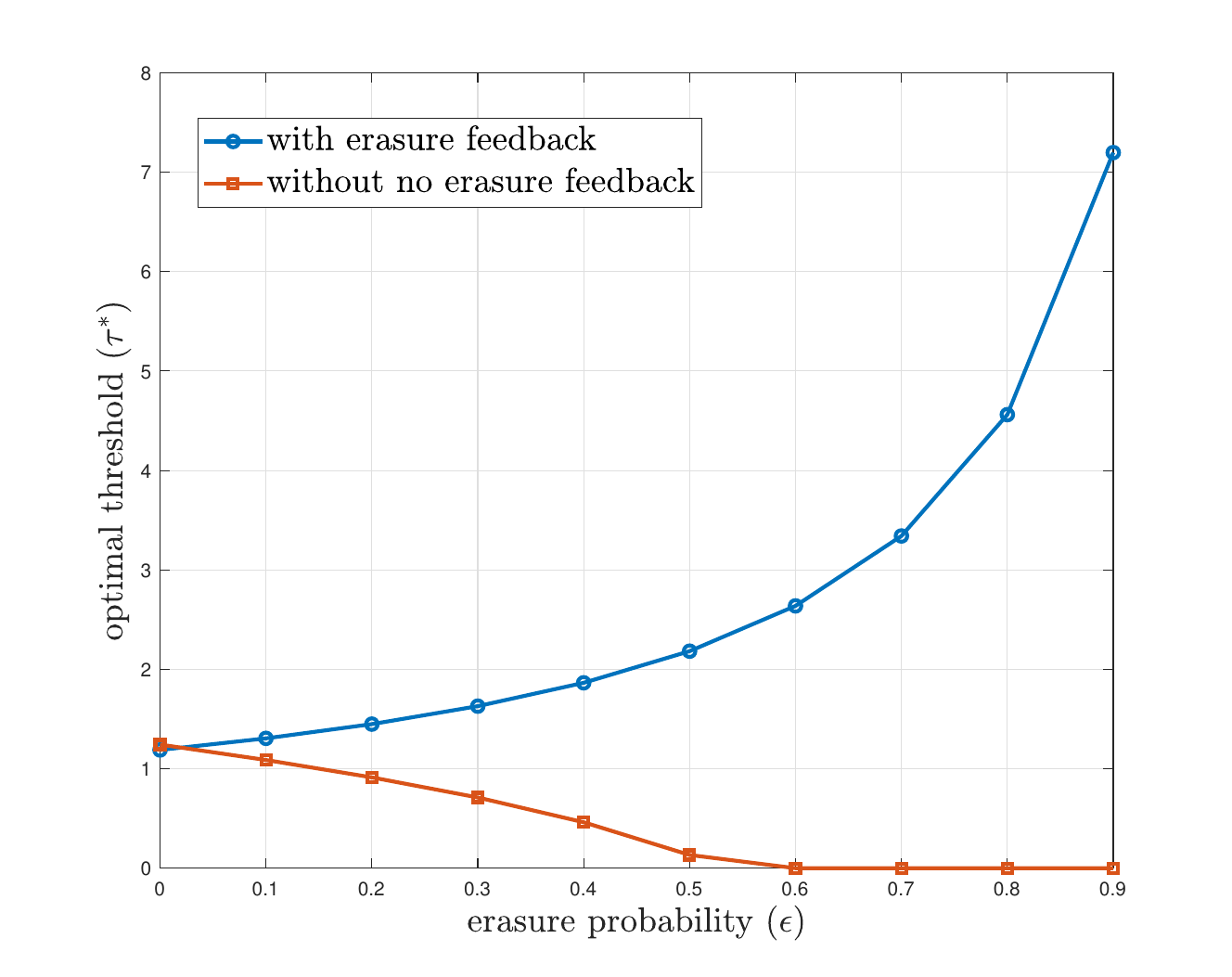}
\caption{Comparison between the optimal threshold behavior versus erasure probability in both cases of the presence (with MAF)/absence (with RR) of erasure status feedback with $f_{\max}=1.5$.}
\label{fig: threshold comparison}
\end{figure}

In Fig.~\ref{fig:threshold_epsilon_NoFB}, we consider the case without erasure status feedback. Our results show that the optimal threshold exhibits the \emph{opposite behavior} compared to the case with feedback (see Remark~\ref{remark_thm2}). Specifically, when the sampling frequency constraint is never binding, i.e., in the case of $f_{\max}=1.5>\mu=1$, the optimal threshold $\tau^*_{RR}$ is {\it decreasing} as the erasure probability increases. Furthermore, when $\epsilon>0.6$, the transmitter employs a zero-wait policy and sends its samples immediately after acquiring them, since $\tau_{RR}^*=0$ in this case. Second, we note that the optimal threshold in the case of $f_{\max}=0.5$ is constant irrespective of the erasure probability. This is due to the fact that the waiting time needed to satisfy the sampling constraint is \emph{independent} of $\epsilon$ in the case of absent erasure status feedback as the waiting appears at the beginning of each transmission round regardless of the erasure events. Finally, we see that for $f_{\max}=0.95$, the optimal threshold is decreasing (following the unconstrained solution of the problem) until it saturates at $\tau_{RR}^*=1$ starting from $\epsilon=0.2$ to satisfy the sampling frequency constraint.

In Fig.~\ref{fig: threshold comparison}, we show the optimal threshold for both cases of erasure status feedback availability at $f_{\max}=1.5$ (non-binding sampling frequency constraint) on the same figure. Fig.~\ref{fig: threshold comparison} shows that both thresholds begin from the same value at $\epsilon=0$. This is due to the fact that both settings are equivalent when there are no erasure events. The optimal thresholds then part their way as the erasure probability increases and behave {\it oppositely} as stated above.

\begin{figure}[t]
\center
\includegraphics[width=0.8\textwidth]{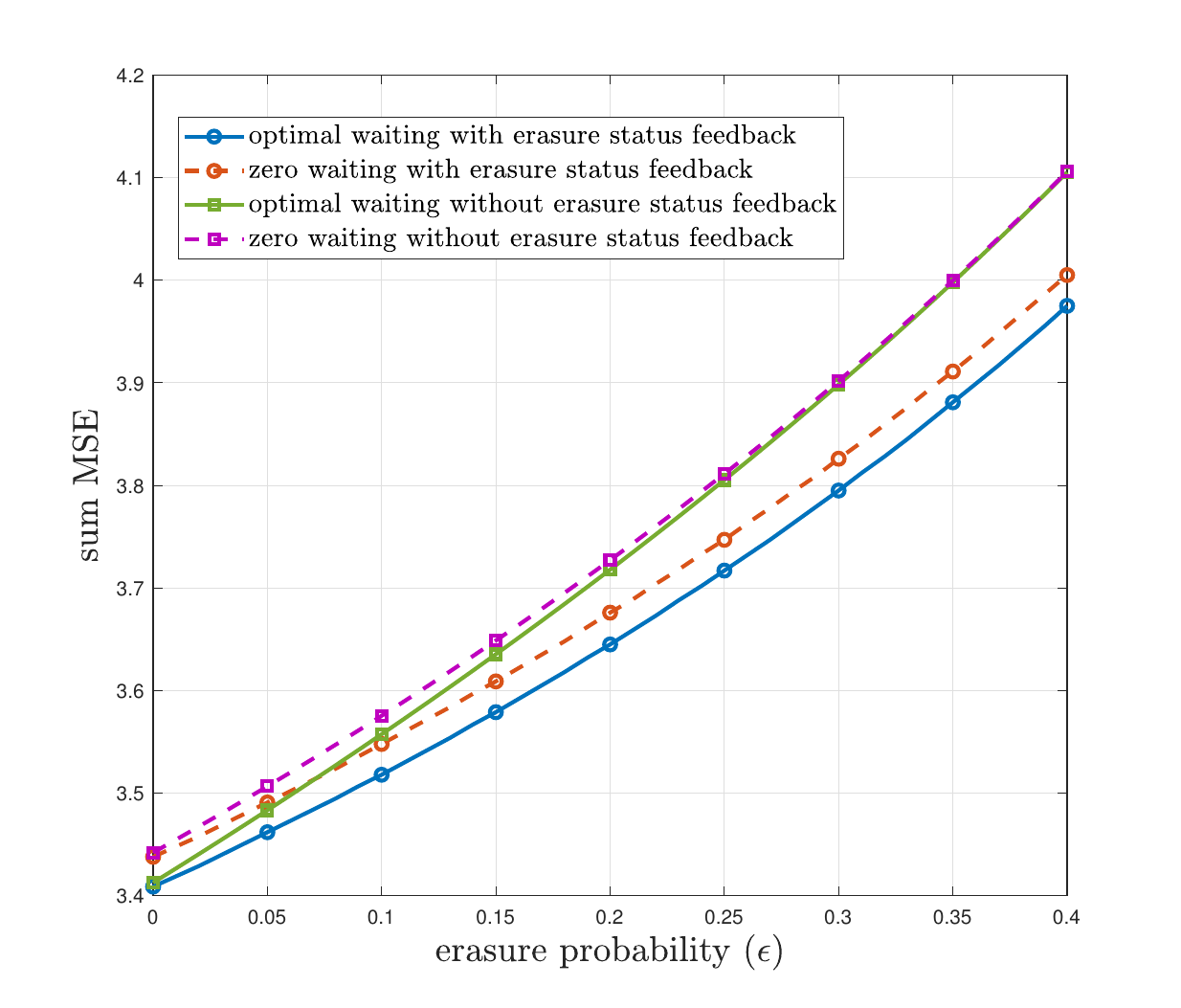}
\caption{Comparison between the sum MSE with optimal waiting (in solid lines) versus the erasure probability in both cases of the presence (with MAF)/absence (with RR) of erasure status feedback with $f_{\max}=1.5$. For each case, we compare the resultant sum MSE with its counterpart if the zero-wait policy is employed (in dotted lines).}
\label{fig: MSE comparison}
\end{figure}

In Fig.~\ref{fig: MSE comparison}, we show the resultant sum MSE versus the erasure probability for both cases of erasure status feedback availability at $f_{\max}=1.5$ (non-binding sampling frequency constraint) on the same figure. We compare the resultant MSE values with their counterparts if the zero-wait policy is employed. Fig.~\ref{fig: MSE comparison} shows that the sum MSE in the presence of erasure status feedback is smaller than that in the absence of erasure status feedback. Furthermore, the sum MSE with optimal waiting for both cases outperforms the zero-wait policy, as expected. In the case of no erasure status feedback, the sum MSE with optimal waiting converges to its counterpart with zero-waiting policy, as the optimal threshold converges to zero as $\epsilon$ increases. Surprisingly, Fig.~\ref{fig: MSE comparison} shows that for small erasure probability (up to $\approx 0.07$), employing \emph{optimal waiting with RR scheduling} outperforms \emph{zero-waiting with MAF scheduling}. I.e., our results show that feedback information can be less impactful if it is not employed to design the optimal sampling instants and use zero-waiting instead of optimal waiting (at least for some range of $\epsilon$).

\subsection{Effect of the Number of Processes}
\begin{figure}[t]
     \centering
     \begin{subfigure}[b]{0.48\textwidth}
         \centering
         \includegraphics[width=\textwidth]{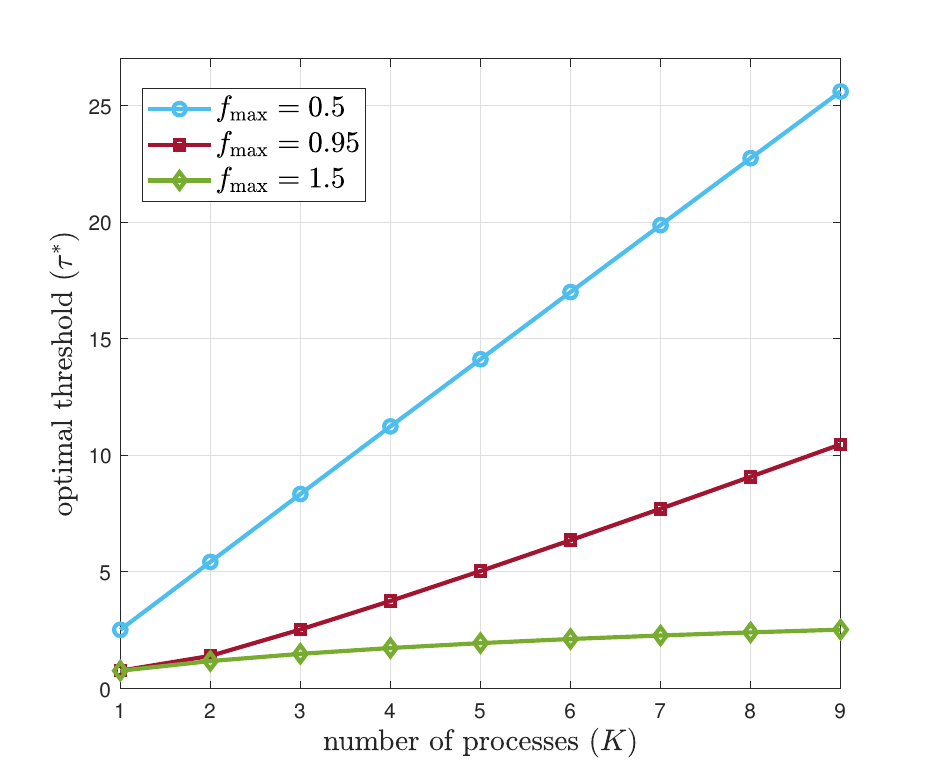}
         \caption{erasure status feedback is present and MAF scheduling.}
         \label{fig:threshold_users}
     \end{subfigure}
     \hfill
     \begin{subfigure}[b]{0.48\textwidth}
         \centering
         \includegraphics[width=\textwidth]{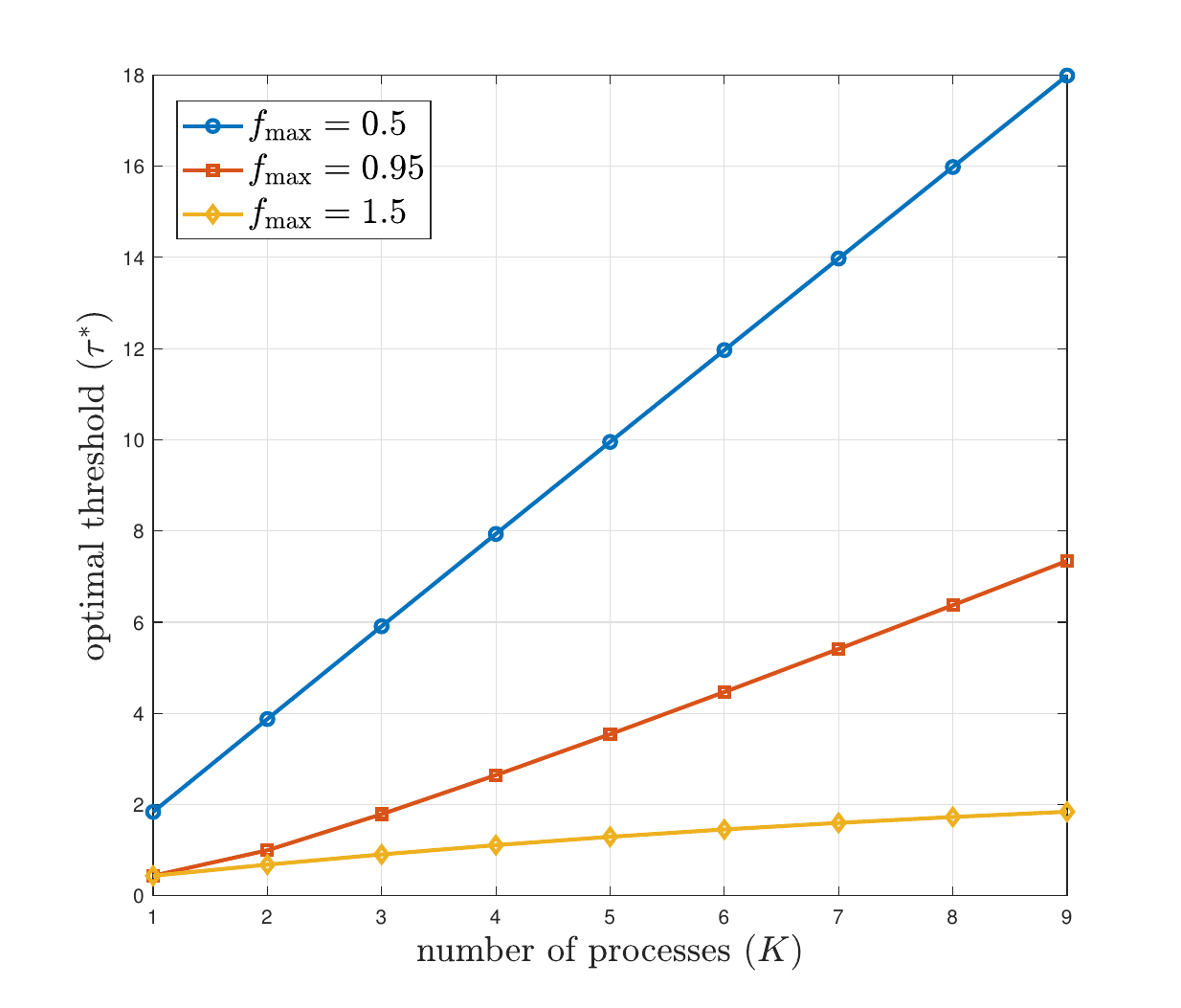}
         \caption{erasure status feedback is absent and RR scheduling.}
         \label{fig:threshold_processes_NoFB}
     \end{subfigure}
     \caption{The optimal threshold ($\tau^*_{\text{MAF}}$ (right) and $\tau^*_{\text{RR}}$ (left)) versus the number of processes ($K$) for different sampling frequency constraints ($f_{\max}$).}
\end{figure}


We now consider a symmetric system with $K$ processes, each having $\sigma_k^2=1$ and $\theta_k=0.5$, for all $k$, communicating through a channel with $\epsilon=0.3$. In Fig.~\ref{fig:threshold_users}, We study the optimal threshold versus the number of processes $K$ in the presence of erasure status feedback. We observe that the long-term average sum MMSE increases with $K$ (as expected, since more processes need to be conveyed through a shared channel). Fig.~\ref{fig:threshold_users} shows that as $K$ increases, the optimal threshold increases as well. The slope of the curve depends on $f_{\max}$. When $f_{\max}=0.5$, the sampling frequency constraint is binding, and $\tau^*$ appears to linearly increase with $K$ with a steeper slope. When $f_{\max}=1.5>\mu=1$, i.e., for an unconstrained version of the problem, the optimal threshold is slowly increasing with $K$. For $f_{\max}=0.95$, the optimal threshold matches the unconstrained solution for $K=1,2$. Nevertheless, when $K>2$, the sampling frequency constraint becomes binding and the linear-like profile of the optimal threshold prevails. 

The exact general patterns hold for the case of the absence of erasure status feedback in Fig.~\ref{fig:threshold_processes_NoFB}. We note, however, that the slopes of curves in Fig.~\ref{fig:threshold_users} are relatively higher than their counterparts in Fig.~\ref{fig:threshold_processes_NoFB}. This is consistent with the results of the previous subsection, where larger waiting (at the same erasure probability) is employed if erasure feedback is present.

\begin{figure}[t]
\center
\includegraphics[width=0.8\textwidth]{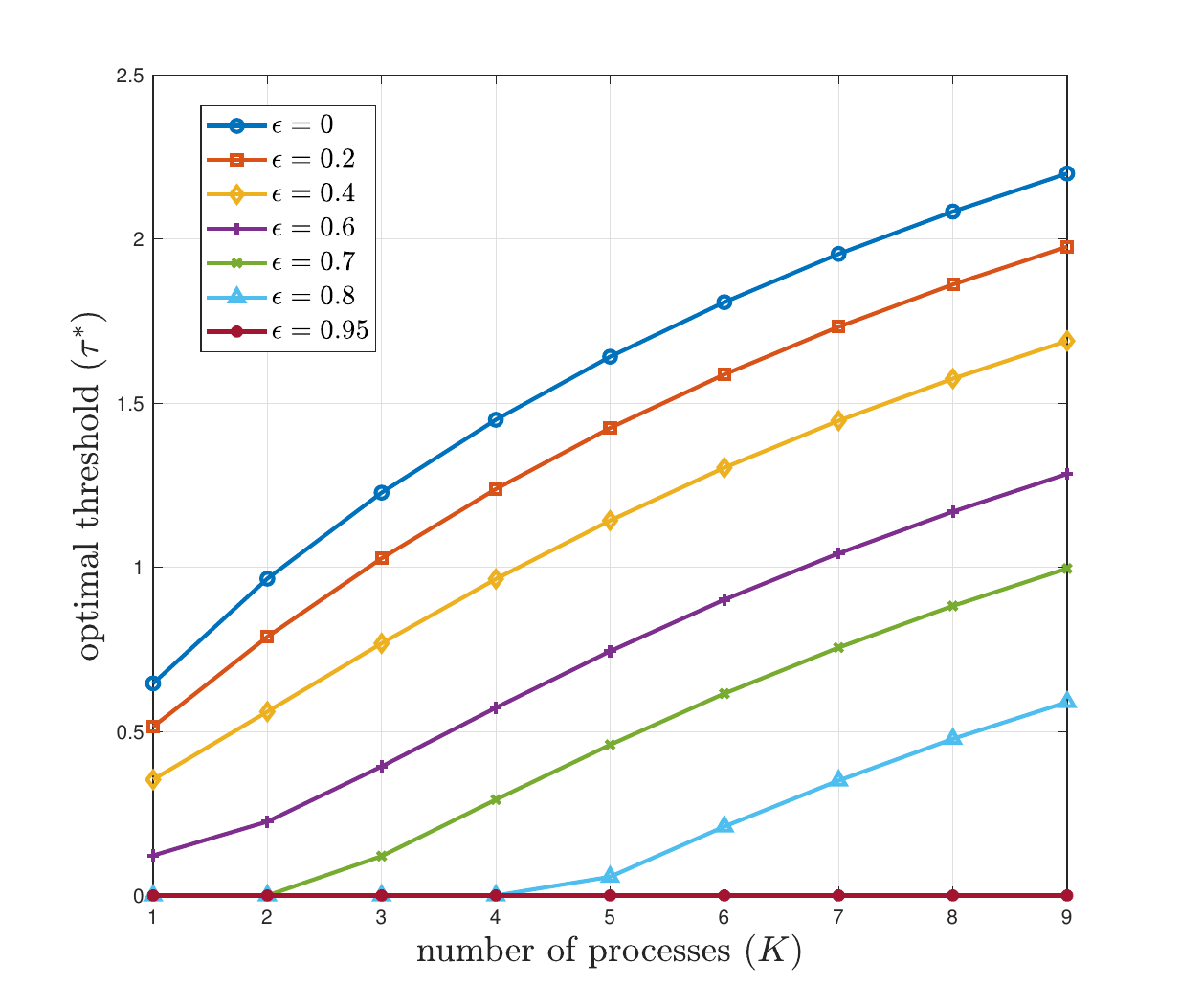}
\caption{The optimal threshold ($\tau^*_{\text{RR}}$) versus the number of processes ($K$) for different values of the erasure probability ($\epsilon$) in the absence of erasure status feedback and RR scheduling.}
\label{fig:threshold_processes_NoFB2}
\end{figure}

Finally, we study how the optimal threshold without feedback, $\tau^*_{\text{RR}}$ behaves with $K$ over a set of different values of $\epsilon$. Interestingly, despite the general increasing trend of the $\tau^*_{\text{RR}}$ versus the number of processes $K$ for fixed $\epsilon$, it is at the same time monotonically decreasing in $\epsilon$ for fixed $K$. This is illustrated in Fig.~\ref{fig:threshold_processes_NoFB2}. This implies that there exists a specific number of processes $\underline{K}$, after which the transmitter starts to wait before each transmission round. This is a contrasting behavior compared to the case of having erasure status feedback, in which increasing $\epsilon$ always increases the $\tau^*_{\text{MAF}}$ for any number of processes.

\subsection{Effect of the Variation Speed of the Processes}
\begin{figure}
     \centering
     \begin{subfigure}[b]{0.48\textwidth}
         \centering
         \includegraphics[width=\textwidth]{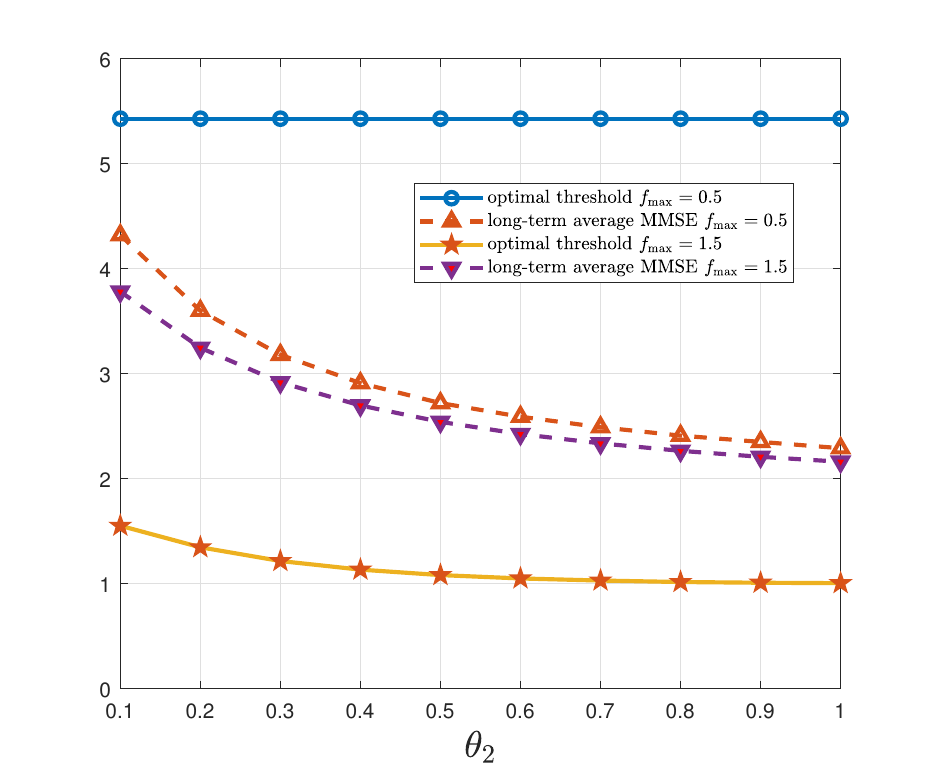}
         \caption{erasure status feedback is present and MAF scheduling.}
         \label{fig:threshold_theta}
     \end{subfigure}
     \hfill
     \begin{subfigure}[b]{0.48\textwidth}
         \centering
         \includegraphics[width=\textwidth]{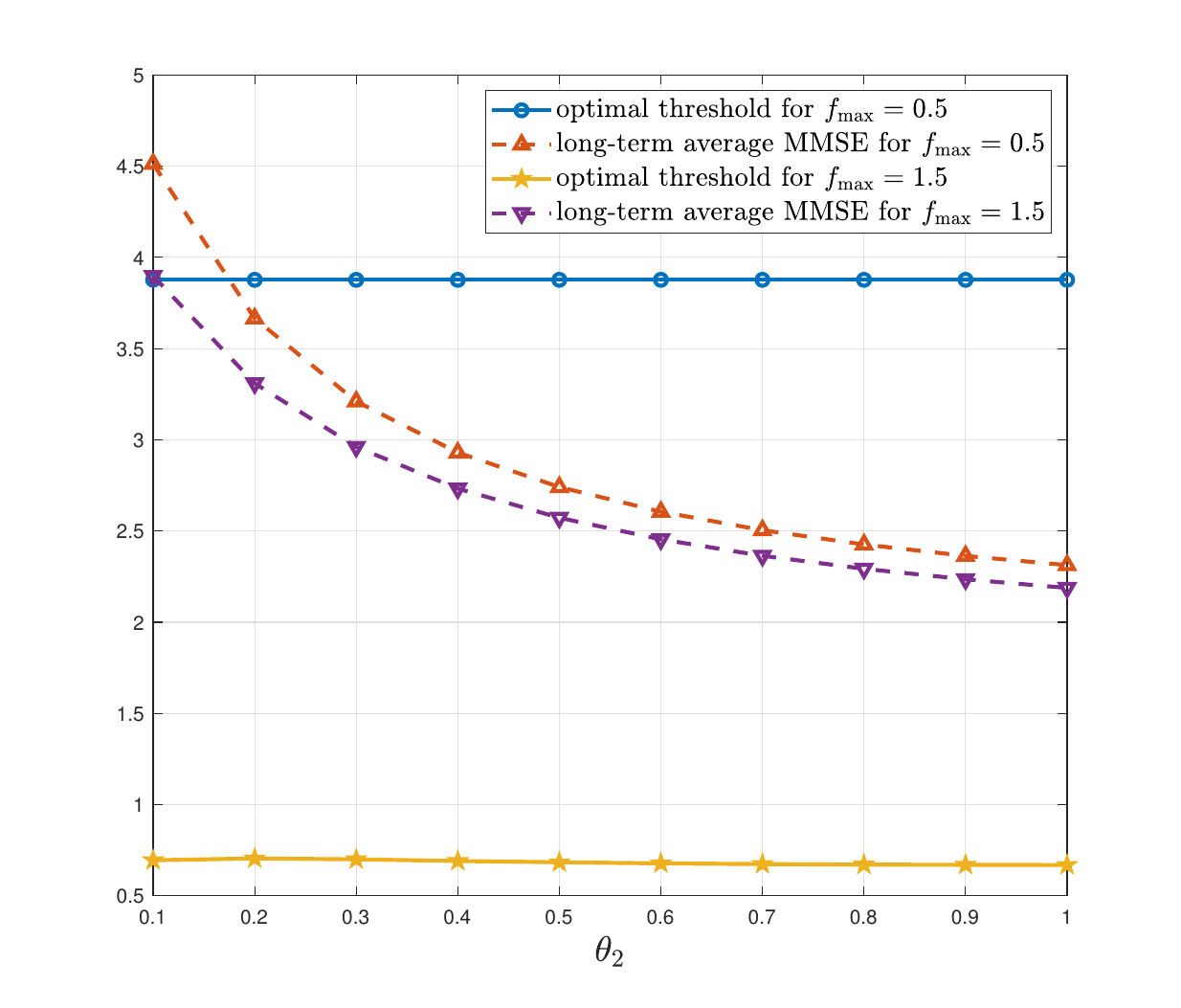}
         \caption{erasure status feedback is absent and RR scheduling.}
         \label{fig:threshold_theta_NoFB}
     \end{subfigure}
     \caption{The optimal threshold ($\tau^*_{\text{MAF}}$ (left) and $\tau^*_{\text{RR}}$ (right)) and the optimal long-term average MMSE versus $\theta_2$ for tracking two processes with different sampling frequency constraints ($f_{\max}$); the first process has fixed parameters $\sigma_1^2=1$ and $\theta_1=1$.}
\end{figure}

We consider a 2-process system with $\sigma_1^2=2$, $\sigma_2^2=1$, and $\theta_1=0.5$. We vary $\theta_2 \in [0.1,1]$ and observe its effect on the optimal threshold and the MMSE for the same service rate $\mu=1$. 

In Fig.~\ref{fig:threshold_theta}, in the case of available erasure status update, we observe that when the sampling frequency constraint is binding, e.g., when $f_{\max}=0.5$, the optimal threshold is independent of $\theta_2$ as the argument of $H^{-1}(\cdot)$ is independent of $\theta_2$. The optimal threshold, however, is a monotonically decreasing function in $\theta_2$ for $f_{\max}=1.5$ as the process becomes faster, and thus the system needs to wait less to track the variations in the process as long as the sampling constraint is inactive. In both cases, the long-term average MMSE is decreasing in $\theta_2$ since the sum of the processes' variances decreases. 

Similar observations can be drawn for the case of the absence of erasure status feedback in Fig.~\ref{fig:threshold_theta_NoFB}. Different from Fig.~\ref{fig:threshold_theta}, it appears that the optimal threshold is changing slightly with respect to $\theta_2$ even for the unconstrained problem (under the considered system parameters). Furthermore, as expected, we can observe that the MMSE function (versus $\theta_2$) for the case of no erasure status feedback is higher than its counterpart if the erasure status feedback is available.



\section{Conclusion}
In this paper, we investigated the problem of estimating $K$ independent OU processes under a total sampling constraint $f_{\max}$, with the goal of identifying the optimal sampling instants such that the long-term average sum MSE is minimized. The acquired samples experience independent erasure events with an erasure probability $\epsilon$. We focused on characterizing the optimal sampling policy in two cases: first, when erasure status feedback is available at the transmitter. In this case, we assume that the transmitter acquires the samples according to the MAF scheduling policy. In the second case, the erasure status feedback is non-existent at the transmitter, and the transmitter employs an RR scheduling policy. We re-formulated both problems in terms of optimizing a stationary waiting policy. In the case of available erasure status feedback, we demonstrated that aggregating waiting at the beginning of the epoch does not hurt the long-term average MSE. In the case of absent erasure status feedback, however, this aggregation needs to be done at the beginning of each RR transmission round.  We showed that the optimal waiting policy is indeed a threshold policy in both settings. We characterized the optimal threshold in terms of $K$, $f_{\max}$, and $\epsilon$. 

Our numerical evaluations and our structural results show an intriguing behavior of the optimal threshold. While, the optimal threshold $\tau^*$ at $\epsilon=0$ is identical for both settings, with increasing $\epsilon$, the optimal threshold for both settings part ways: it \emph{increases} for the case of \emph{available} erasure status feedback, and \emph{decreases} when the erasure status feedback is \emph{non-existent}. Furthermore, we show that for both settings, the optimal threshold is an increasing function of the number of processes $K$. Finally, we show the effect of the variation speed of the process on the long-term average sum MSE.

Future directions of this work may include investigating fully-observable processes (OU or otherwise), different age-dependent penalties other than the sum MSE penalty, the behavior of the waiting policy for generalized statistical models for the service queue other than the exponential distribution, exogenous sampling with and without preemption rather than the generate-at-will model, and signal-dependent sampling rather than signal-independent sampling.









\bibliographystyle{unsrt}
\bibliography{reference}
\end{document}